# Breaking the angular dispersion limit in thin film optics by ultra-strong light-matter coupling

Andreas Mischok[1,2,*], Bernhard Siegmund[3], Florian Le Roux[1], Sabina Hillebrandt[1,2], Koen Vandewal[3], Malte C. Gather[1,2,*]

[1]*Humboldt Centre for Nano- and Biophotonics, Department of Chemistry, University of Cologne, Greinstr. 4-6, 50939 Köln, Germany*

[2]*School of Physics and Astronomy, University of St Andrews, North Haugh, St Andrews KY16 9SS, United Kingdom*

[3] *UHasselt, Institute for Materials Research (IMO-IMOMEC), Agoralaan, Diepenbeek, 3590 Belgium*

**contact: andreas.mischok@uni-koeln.de, malte.gather@uni-koeln.de*

**Abstract**

**Thin film interference is integral to modern photonics and optoelectronics, e.g. allowing for precise design of high performance optical filters, efficiency enhancements in photovoltaics and light-emitting devices, as well as the realization of microlasers and high-performance photodetectors. However, interference inevitably leads to a change of spectral characteristics with angle, which is generally undesired and can limit the usefulness of thin-film coatings and devices. Here, we introduce a strategy to overcome this fundamental limit in optics by utilizing and tuning the exciton-polariton modes arising in ultra-strongly coupled microcavities. We demonstrate optical filters with narrow pass bands that shift by less than their half width (<15 nm) even at extreme angles. Our filters cover the entire visible range and surpass comparable metal-dielectric-metal filters in all relevant metrics. By expanding this strategy to strong coupling with the photonic sidebands of dielectric multilayer stacks, we also obtain filters with high extinction ratios and up to 98% peak transmission. Based on these findings, we realize ultrathin and flexible narrowband filter films, monolithically integrate our filters with organic photodiodes, and demonstrate polarization-sensitive polariton filters. These results illustrate how strong coupling provides additional degrees of freedom in thin film optics that will enable a multitude of exciting new applications in micro-optics, sensing, and biophotonics.**

Two reflective surfaces form a basic *Fabry-Pérot* microcavity with a narrowband spectral response at a resonance wavelength $\lambda_{\mathrm{res},0}$ if the distance $d$ between the surfaces equals approximately $m\lambda_{\mathrm{res},0}/2n$, where $m$ is an integer and $n$ is the refractive index of the material between the cavity mirrors. If light impinges on this structure under an oblique angle $\theta$, the photon wavevector acquires an additional in-plane component, and thus the resonance shifts to a shorter wavelength, i.e. $\lambda_{\mathrm{res}}(\theta) = \lambda_{\mathrm{res},0}\sqrt{1-(\sin\theta/n)^2}$. This angular dispersion is considered to be a fundamental property of all thin optical films and must be taken into account during device design[1–8]. Thin film angular dispersion is sometimes utilized to fine-tune the spectral response of a device by tilting, but in practice thin-film angular dispersion is generally undesired for a number of reasons: (1) the presence of angular dispersion requires precise alignment and makes optical systems prone to drift over time; (2) in display applications, it can lead to changes in perceived colour for different viewing angles; (3) if light with a distribution of angular components (as is the case for most light sources) passes through a thin film interference filter, the wavelength selectivity of the filter is compromised and the transmitted light is spectrally broadened in an uncontrolled way; (4) finally, the transmitted line-shape of optical components is often severely distorted at large angles, in part because the response of thin optical films depends on the polarisation of the incoming light (polarisation-splitting)[1,9].

So far, the main remedy for angular dispersion has been the use of high refractive index materials[10]. However, this approach is technically challenging, often introduces optical losses, and has limited efficacy. More elaborate designs for managing angular dispersion include dielectric and plasmonic nanostructures[11–17], lossy *Fabry-Pérot* cavities[16,18–23], and multilayer structures to compensate phase[24–26] or induce Fano resonances[27]. All these strategies however typically suffer from either high losses[28], low optical quality, or poor angular performance. Compared to conventional thin-film coatings, nanostructure-based filters also require a significant fabrication effort.

Here, we demonstrate how the fundamental angular dispersion limit in thin-film optics can be overcome by making use of the exciton-polariton dispersion in strongly and ultra-strongly coupled microcavities[29–32]. Exciton-polaritons are quasiparticles that form through coherent interaction between an optical mode and an excitonic material resonance. The strongly coupled resonances split into a low energy (lower polariton) and high energy (upper polariton) branch once their coupling strength exceeds the losses of the individual resonances. The ultra-strong coupling regime is reached when this splitting becomes significant (~20%) relative to the bare exciton energy[30]; in the visible range, such splittings can equate to $> 100$ nm in wavelength[33]. The characteristic shape of the polariton dispersion in optical microcavities has been studied extensively, and the distinctive anti-crossing between the lower and upper branch is often regarded as proof for the presence of strong coupling. It has been suggested that polariton dispersion can be exploited to improve colour stability in information displays[33–35]; however, strong coupling has not been explored systematically for managing dispersion in thin-film optics. Strongly and ultra-strongly coupled exciton-polaritons have instead attracted great interest as a platform for studying light-matter interaction[32,36,37], with applications expected e.g. in polariton lasing[38–41], polariton chemistry[42], light emission[35,43–46], photodetection[47–49] and quantum information processing[50,51]. Organic semiconducting materials are of particular interest for these efforts as they can reach the ultra-strong coupling regime at room temperature[33] due to their exceptionally high oscillator strengths, broad absorption bands and large exciton binding energies. In addition, organic materials offer great spectral tunability and simple, low-cost and scalable fabrication.

We show here that by carefully adjusting the coupling strength and energetic offset (“detuning”) between an optical and an excitonic resonance, one obtains polaritonic dispersions that trace the angle-independence of organic excitons yet retain the spectral purity of an optical resonance. We demonstrate this concept by realizing narrowband transmission filters based on organic microcavities that operate in the strong and ultra-strong coupling regime. These filters exhibit spectral responses that are almost angle- and polarization-independent, with shifts in peak transmission wavelength over the full range of incident angles smaller than their respective linewidths. Utilizing different organic materials, we show the applicability of the concept throughout the visible spectral range and into the near infrared. As a further generalization, we introduce organic materials into multilayer dielectric stacks, which induces sideband coupling. This approach affords optical coatings that offer extinction ratios comparable to state-of-the-art dielectric filters, while severely suppressing the angle-dependence of their transmission spectra. Finally, to showcase the application potential of polariton-based filters, we explore three different scenarios: mechanically flexible filters with spectral characteristics that are robust to bending; integration in a sensitive, narrowband, angle-independent photodetector; and polarizing polariton filters utilizing anisotropic strong coupling.

## Results

Figure 1**a** schematically shows a transmissive narrowband *Fabry-Pérot* microcavity. If the material inside the cavity is transparent or weakly absorbing, the system is in the regime of weak light-matter coupling and the cavity exhibits a mostly parabolic dispersion with a substantial shift in resonance wavelength for different angles of incidence (Fig. 1**b**). Introducing a material with a strong excitonic resonance into the cavity gives rise to strong light-matter coupling and leads to a splitting of the original parabolic dispersion into two polariton branches: The lower polariton branch (LPB) is red-shifted with respect to both the photon and exciton resonances, shows approximately photonic parabolic dispersion at low angles, and flattens out to a more exciton-like dispersion at larger angles. The second, higher energy, upper polariton branch (UPB) shows the inverse behaviour. Utilizing strong coupling adds additional design parameters not available in conventional thin film optics design – namely the coupling strength and the cavity detuning. The coupling strength $\hbar\Omega_{\mathrm{R}}/2$ describes the interaction between photons and excitons and thus the energetic (*Rabi*-)splitting ($\hbar\Omega_{\mathrm{R}}$) between LPB and UPB. The detuning $\delta$ is the difference between the energies of the bare photon resonance at normal incidence ($E_P$) and bare exciton resonance ($E_X$). We find that tuning the microcavity to a relatively large absolute value of $\hbar\Omega_{\mathrm{R}}$ and a moderate positive detuning $\delta$ in the range of 100 meV - 200 meV results in a mostly flat LPB and thus largely avoid shifts in resonance wavelength with angle (Fig. 1**d**). These conditions also keep a high absolute transmission; even stronger positive detuning results in a loss of peak transmission while going to negative detuning makes the LPB more photonic and thus introduces strong angular dispersion again (Supporting Information, Fig. S13). Using a large $\hbar\Omega_{\mathrm{R}}$ helps in two ways: first, it increases the separation between LPB and UPB, making it possible to filter out the UPB if desired (see Supporting Information, Fig. S2). Second and more importantly, a large coupling strength (i.e., going into the ultrastrong coupling regime) creates a larger separation between the LPB and the bare exciton absorption, helping to avoid parasitic absorption from uncoupled excitons while keeping the overall dispersion flat (further details, see Supplementary Discussion 1 and 3).

Figure 1**e** illustrates this strategy by comparing transfer-matrix (TM) calculations of optical transmission versus angle and wavelength for different microcavities comprising a layer of either transparent $SiO_2$ or the organic absorber C545T sandwiched between two semi-transparent silver mirrors. The difference in the angle-dependent behaviour is immediately obvious; the presence of ultra-strong coupling in the cavity with the organic absorber leads to a large splitting ($\hbar\Omega_R \geq 1\mathrm{eV}$) and –in combination with the positive detuning– results in a near complete flattening of the dispersion compared to the uncoupled cavity based on $SiO_2$. (See Supplementary Information Table S2 for data on uncoupled cavity based on high index $Ta_2O_5$).

Our concept is not limited to simple metallic cavities. Recently, it has been shown that excitonic resonances also couple with the Bragg sidebands of distributed Bragg reflector (DBR) cavities made of alternating high and low refractive index layers[52]. We utilize this behaviour to design optical coatings with high extinction and sharp edge characteristics. TM simulations of a single DBR and a strongly coupled cavity containing a layer of C545T sandwiched by two similar DBRs illustrate how the Bragg sidebands of the mirror couple to the C545T exciton in a similar manner as the cavity mode in the previous example (Fig. 1**f**). However, in contrast to the metallic cavities, they now form multiple Bragg-polaritons[52,53], leading to a series of lower polariton branches, each with reduced angular dispersion.

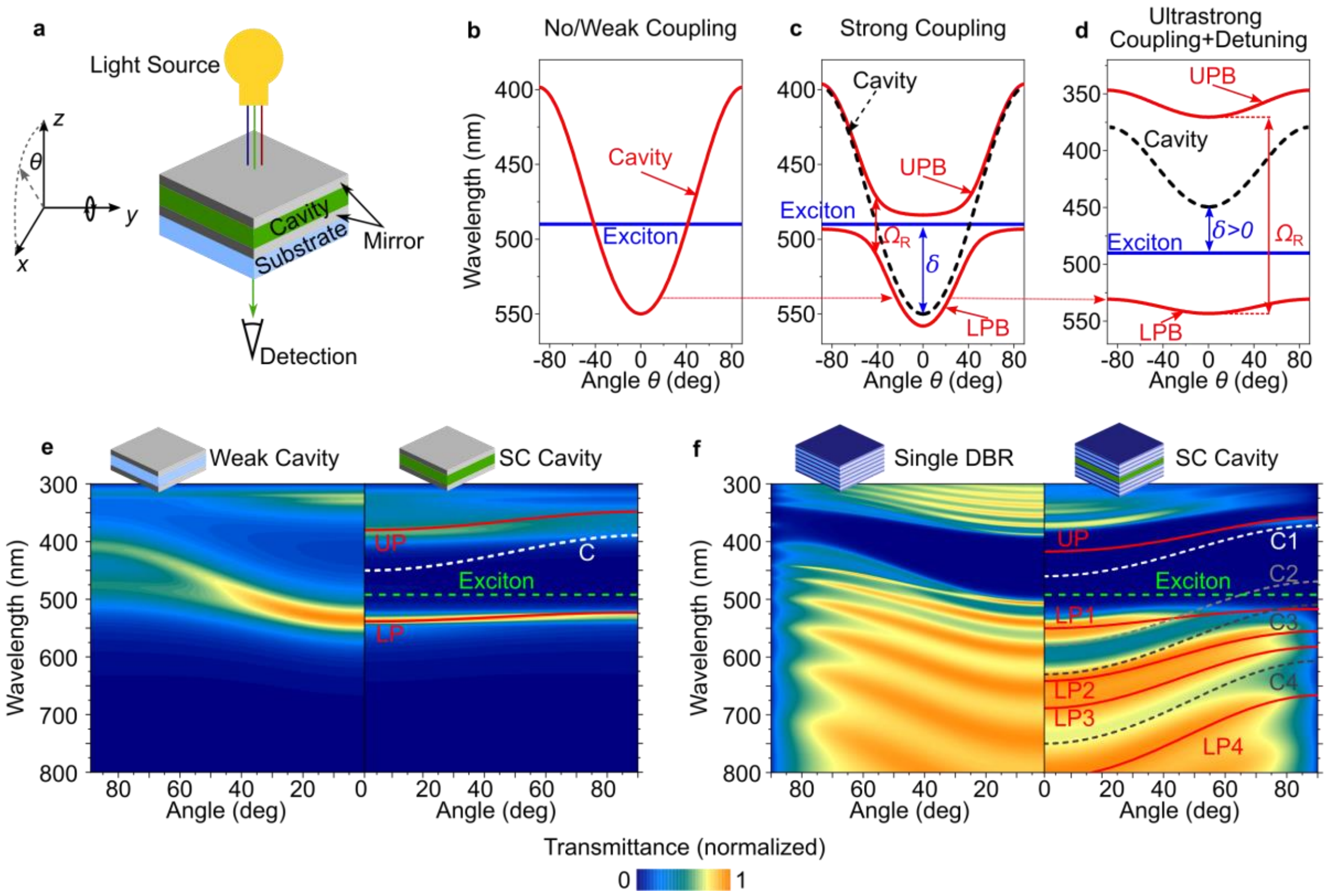

**Figure 1: General concept of polariton dispersion management and optical modelling of polariton filters**. **a** Sketch of cavity-based transmission filter between light source and detector, with a variable angle of incidence $\theta$. **b-d** Transfer matrix simulation of the angular dispersion of an uncoupled or weakly coupled cavity with intracavity refractive index $n \approx 1.45$ (**b**), a strongly coupled cavity with Rabi-splitting $\Omega_R \approx 0.2\,\mathrm{eV}$, negative detuning $\delta \approx -0.2\,\mathrm{eV}$ and $n \approx 1.85$ (**c**), and an ultra-strongly coupled cavity with $\hbar\Omega_R \approx 1\,\mathrm{eV}$ and positive detuning of $\delta \approx 0.2\,\mathrm{eV}$ (**d**). Strong coupling leads to formation of upper and lower polariton branches (UPB/LPB) which avoid the crossing point of cavity photon (dashed line) and material exciton (solid blue line). In each panel, the transmitted modes are indicated as solid red lines. While the uncoupled cavity shows a characteristic and large angular dependence, strong coupling can dramatically reduce this dependence; under ultra-strong coupling conditions in particular, the angular dispersion can be almost negligible. **e** Transfer matrix simulation of angle-resolved transmission of a metal-dielectric-metal weakly coupled ("Weak") cavity (left) and a strongly coupled (SC) cavity (right). Angular dispersion leads to a significant blueshift and polarization splitting of the narrowband transmission for the weak cavity. For the SC cavity, the resonance is further narrowed and shows almost no blueshift and polarization splitting at large angles. Lines represent a coupled-oscillator model of polariton modes (red lines) as well as the bare exciton (green dashed) and cavity (white dashed) resonances. **f** Simulated transmission of a distributed Bragg reflector (DBR, left)

and a blue-shifted strongly coupled DBR cavity (right). For the latter, the DBR sidebands couple to the excitonic resonance, leading to an angle-independent stopband between 400 nm and 500 nm.

To demonstrate the proposed polariton filter design experimentally, we choose several common organic semiconductors that show strong excitonic absorption throughout the visible spectrum, starting with the organic dye C545T already used in the above simulations (details in Supplementary Information, Discussion 1). Figure 2**a-f** compares the measured, angle-resolved transmittance spectra of a conventional Ag-$SiO_2$-Ag microcavity filter to a Ag-C545T-Ag polariton filter. The C545T-based filter exhibits clear polariton formation with high transmission in the UPB and LPB, respectively (compare calculated polariton modes in Fig. 1**e**) and surpasses the conventional filter in all metrics: (1) it shows a sharper linewidth (27 nm vs 56 nm full width at half maximum (FWHM)), (2) a higher rejection on the blue side (minimum transmission 0.9% vs 5.7%), (3) a better spectral tolerance to thickness variations (Supporting Information, Figure S5), (4) a drastically improved angular dispersion, and (5) higher transmission at large angles of incidence (Supporting Information, Figure S 11). While the conventional filter exhibits a spectral shift of >100 nm and a drastic increase in linewidth due to polarization splitting, the polariton filter shows a polarization-independent spectral shift of <15 nm and a stable linewidth even at extreme angles. For an angle range of <70°, its spectral shift is <10 nm, while for the conventional filter, it is ≈100 nm. The enhanced angular stability is strikingly obvious when looking at the filter at different angles by eye (Fig. 2**c**,**f**). If transmission in the wavelength range of the UPB is undesired, it can be blocked by a complementary absorption layer (Supporting Information, Figure S2).

While such basic cavity filters are easy to design, their performance and versatility, e.g. in terms of controlling lineshape, are limited. To tune the spectral response and enhance transmission, additional dielectric layers were added to the metal-metal polariton filters, with a final design of $Ta_2O_5$(80 nm)-$SiO_2$(62 nm)-$Ta_2O_5$(58 nm)-Ag(25 nm)-C545T(100 nm)-Ag(25 nm)-$SiO_2$(176 nm)-$Ta_2O_5$(71 nm). This allowed us to shape the electric field in the cavity and thus reduce losses caused by parasitic absorption in the metal. Figure 2 **g-i** showcases such a dielectric-supported filter that offers a broader passband (FWHM ≈ 85 nm), exhibits a peak transmission of 80% and shows a virtually unchanged transmission lineshape up to a 40° angle of incidence. In this design, there is a trade-off between peak transmission and linewidth, as the additional dielectric layers reduce the reflectivity of the metal layers and place them at a field minimum, which in turn slightly reduces cavity quality. This effect could be remedied by the addition of further layers to boost the overall reflectivity again.

Next, we performed a series of transfer matrix simulations to investigate if the absorption of the organic materials introduced to obtain strong light-matter coupling limits the peak transmission achievable in polariton filters. We find that for both conventional metal-metal microcavity filters and metal-metal polariton filters, the transmission performance is similar. For both architectures performance is ultimately limited by residual absorption in the metallic layers and by reflection at the outer air-metal interface, with absorption in the organic materials not representing a significant loss pathway (absorption at peak <3%, Supplementary Discussion 3 and Figures S11-S16).

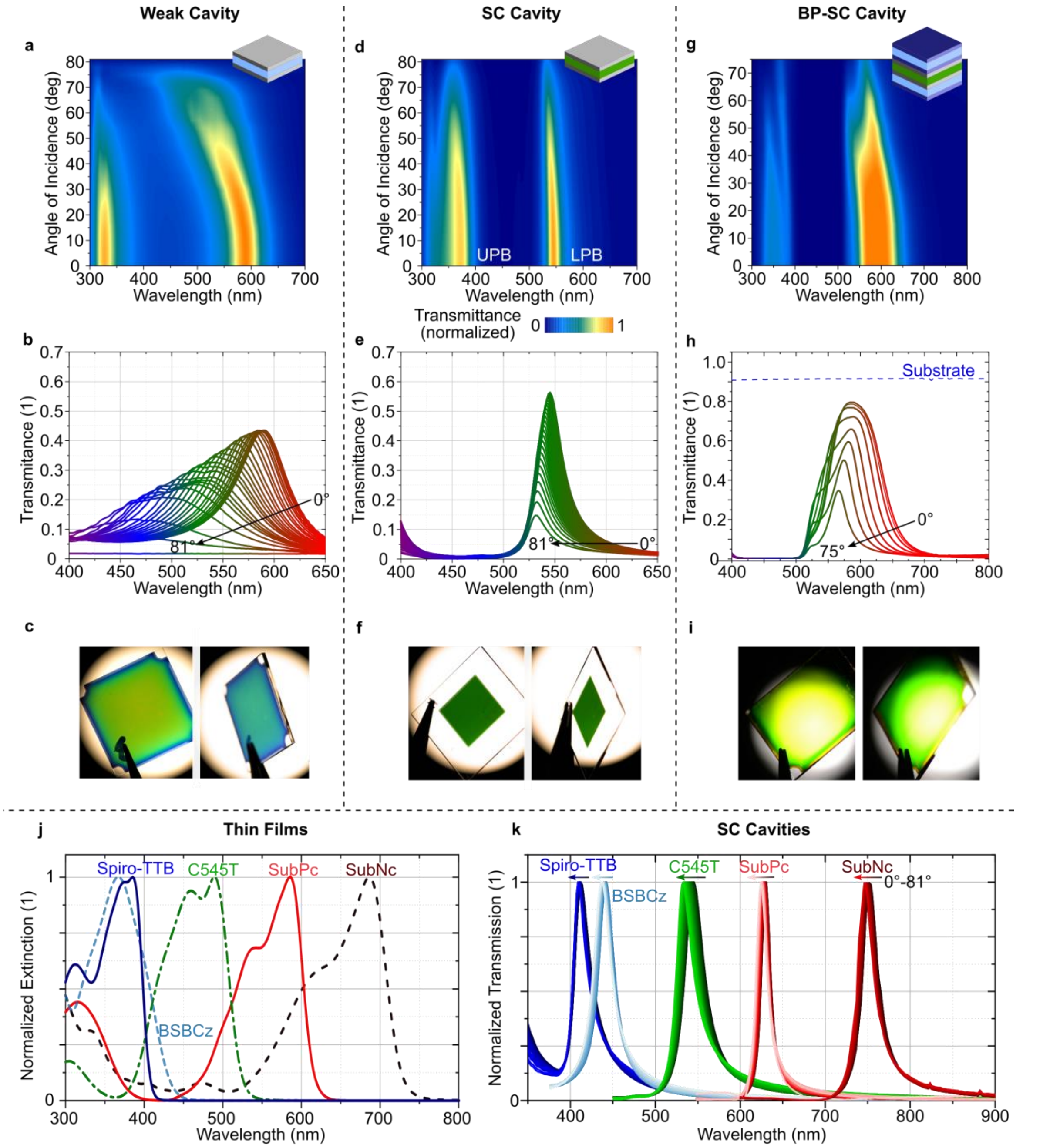

**Figure 2: Angle-resolved transmission of conventional and polariton-based metal cavity filters. a,b** Measured angle-resolved transmittance of a conventional, weak Ag(25 nm)-$SiO_2$(140 nm)-Ag(25 nm) cavity as false colour map (**a**) and line plots (**b**). Due to dispersion, the position of the transmitted mode strongly depends on the angle of incident light, shifting by more than 100 nm and exhibiting strong polarization splitting at large angles. **c** Photographs of the weak cavity under normal and oblique angles. **d,e** Measured angle-resolved transmittance of an ultrastrongly coupled Ag(25 nm)-C545T(80 nm)-Ag(25 nm) cavity as false colour map (**d**) and line plots (**e**). Ultra-strong coupling drastically reduces dispersion, resulting in a mode shift of <15nm over the entire measured angular range. **f** Photographs of the polariton filter under normal and oblique angles. **g,h** Measured angle-resolved transmittance of a bandpass filter (BP-SC) based on an ultrastrongly coupled Ag(25 nm)-C545T(100 nm)-Ag(25 nm) cavity with additional dielectric layers as false colour map (**g**) and line plots (**h**). Additional high ($Ta_2O_5$) and low ($SiO_2$) index dielectric layers are utilized to tune the transmission spectrum for a broader bandpass and higher peak transmittance (up to 80%), with high stability for angles up to 40°. Dashed line in (**h**) indicates transmission of the bare substrate. **i** Photographs of the bandpass polariton filter under normal and oblique angles. **j** Extinction spectra of five different organic

materials used in this work. **k** Normalized transmission spectra of polariton filters based on the materials in **j**. Each set of spectra shows the transmission of the LPB for angles of incidence from 0° to 81°, with an interval of 3°. By selecting suitable absorbers, polariton filters with narrowband angle-independent characteristics can be created across the whole visible spectrum, e.g. with Spiro-TTB (415 nm, blue), BSBCz (440 nm, light blue), C545T (540 nm, green), $Cl_6$SubPc (630 nm, light red) and SubNc (750 nm, dark red).

The chemical diversity of organic materials enables the application of the polariton filter concept across a wide spectral range. Figure 2**j** depicts the normalized extinction of the materials Spiro-TTB[54], BSBCz[55,56], C545T[57], $Cl_6$SubPc, and SubNc[58] (full chemical names in methods). We found that in general, materials with strong absorption and steep absorption onset are well suited to form exciton-polaritons with suitable properties to realize narrowband, angle-stable filters, while photoluminescence efficiency does not play a role. Therefore, different classes of organic materials, from fluorescent emitters (BSBCz, C545T) to photovoltaic absorbers ($Cl_6$SubPc, SubNc) and charge transport layers (Spiro-TTB), can be utilized to design such filters. As an example, we realized filters with transmission lines between 400 nm and 800 nm, with 35 nm thick Ag mirrors sandwiching the organic material (Fig. 2**k**). The transmission spectra of the different filters are well separated at all angles, as required for a multitude of applications in spectroscopy, hyperspectral imaging, microscopy, and coloration. A detailed analysis of the metal-organic-metal filters, including full spectra, and variations in organic and metal layer thickness, is presented in the Supplementary Information, Discussion 1 and Figures S1-S6.

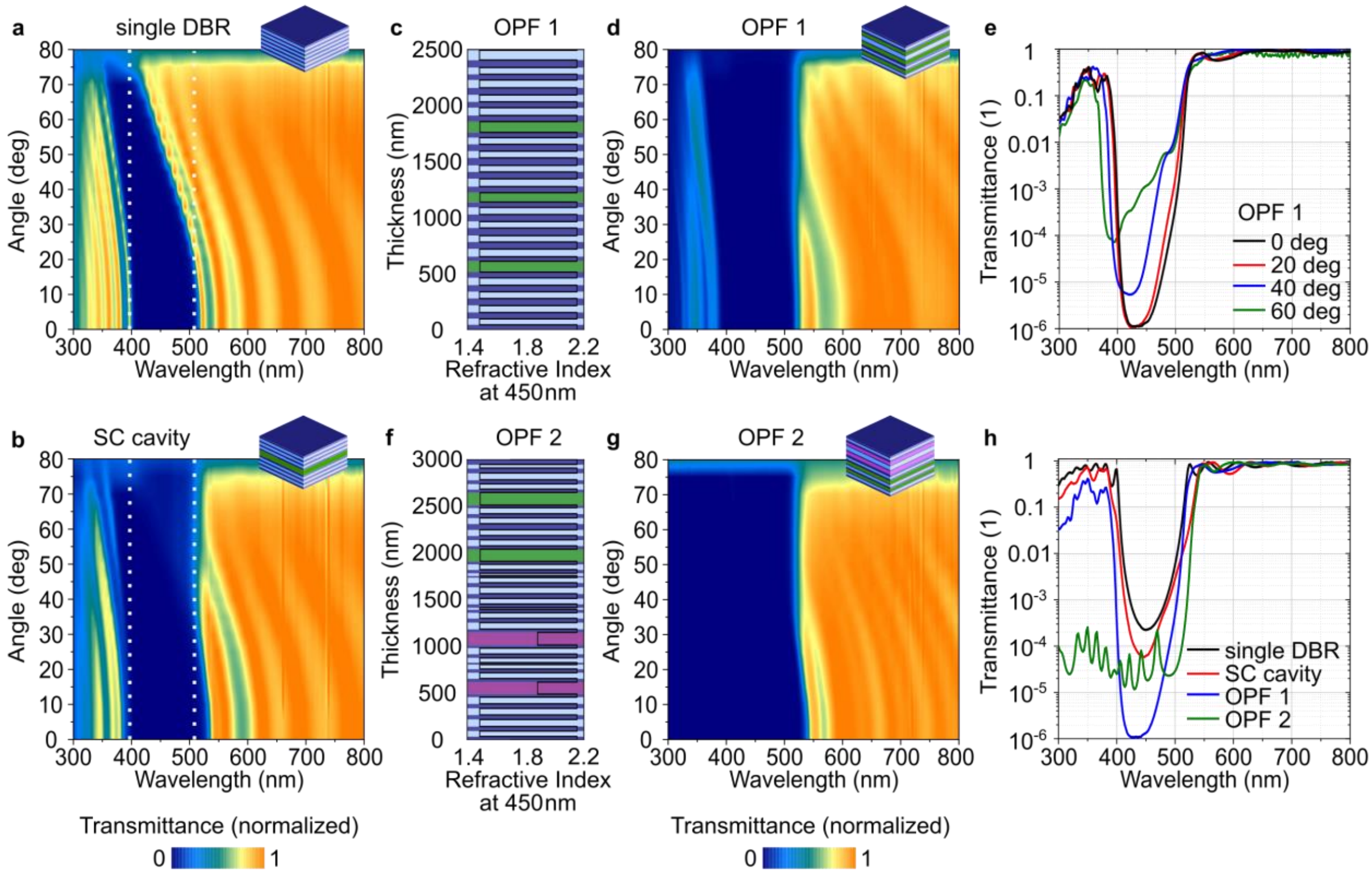

**Figure 3: DBR based polariton filters with high extinction. a-d** Experimental angle-resolved transmission spectra of a single DBR with 21 alternating layers of $Ta_2O_5$ and $SiO_2$ (**a**), and a strongly coupled cavity with two 11-layer DBRs sandwiching an organic C545T core layer (**b**). As a guide to the eye, dashed lines indicate the position of the stopband at 0°. **c** Layer stack of a computationally optimized filter design (Optimized Filter 1, OPF 1) comprising three layers of C545T (green), embedded in $Ta_2O_5$ (dark blue) and $SiO_2$ (light blue). **d** Experimental angle-resolved transmission spectra of OPF 1 in false colour scale. **e** Transmission spectra of OPF 1 at different angles on logarithmic scale. **f** Layer

stack of a computationally optimized broadband filter design (Optimized Filter 2, OPF 2) comprising two Spiro-TTB layers (purple) and two C545T layers (green). **g** Experimental angle-resolved transmission spectra of OPF 2 on false colour scale. While the stopband of the conventional DBR shows a strong angular dependence, the optimized polariton long-pass filters show largely angle-independent stopbands with high extinction and high transmission (>80%) at long wavelengths. **h** Transmission spectra of filters from **a-g** at normal incidence on logarithmic scale.

As metal films introduce parasitic absorption, metal mirror-based filters are limited in terms of the maximum transmission and extinction they can achieve. Most applications requiring high optical quality, such as high passband transmission and high optical density in the stopband, therefore employ dielectric multilayer stacks, which comprise alternating films of low and high-refractive index materials. As already shown by the optical modelling above, embedding a material with a strong excitonic resonance in a cavity formed by dielectric mirrors can induce strong coupling with the sidebands of the mirrors. To further test this strategy, we fabricated and compared a single 21 layer DBR and a 2×11-layer DBR-based cavity with a core layer of C545T in the strong coupling regime (Fig. 3**a,b**). In contrast to previous demonstrations of strong coupling in DBR cavities, we strongly blue-shifted the design wavelength such that the red edge of the stopband coincides with the exciton resonance. The resulting side-band coupling flattens the DBR response at large angles and leads to a near continuous stopband between 400 nm and 500 nm up to angles of incidence of 70°.

To further improve filter performance, we computationally optimized the individual layer thicknesses for angle-independent transmission using particle swarm optimization[59,60] followed by Levenberg-Marquardt optimization[61]. For Optimized Filter 1 (OPF 1, Fig. 3**c,d**), we introduced three 100 nm thick C545T layers to facilitate strong coupling and optimized the surrounding dielectric layers to achieve a high transmission >93% above 530 nm and a high rejection in the stopband, while limiting the total number of layers to 41, which is still low compared to many commercial filters. While fabrication-related uncertainties in the thicknesses of the individual layers led to some deviations from the predicted behaviour, we were nevertheless able to experimentally demonstrate a filter with a largely angle-independent stopband position that reaches an optical density (OD) of 6 and retains an OD of 4-5 even at large angles of incidence (Fig. 3**e**, compare simulations and conventional filter in Supplementary Fig. S7 and performance analysis in Supplementary Discussion 3). Furthermore, through application of a backside anti-reflection-coating (ARC), the passband transmission can be increased to >98% (Supplementary Fig. S15).

In a second optimization, we aimed to reduce the transmission of the blue sideband to create a longpass filter. To achieve this, we introduce a second organic material, Spiro-TTB, integrated in a blue-shifted DBR located on top of the C545T-based filter. Computational optimization led to a design comprising 2×2 organic cavities and a total of 55 layers. Experimental realization of this design yielded a longpass filter (OPF2) with a broad stopband OD >3 across a large range of angles up to 70° (Fig. 3**f,g**). Figure 3**h** compares the measured transmission of the four presented filters on logarithmic scale. (See Supplementary Information, Discussion 2 and Figures S7-S10 for further analysis.) These results illustrate how side-band coupling between highly absorbing materials and a dielectric multilayer stack can lead to a multitude of novel filter designs with drastically reduced angular dispersion. The performance of the DBR-based polariton filters discussed above is summarized in Table 1.

**Table 1. Performance of DBR-based polariton filters.**

| Filter-Type | Max T | Max T w/ ARC | Max OD | Δ*λ* at 45° | Δ*λ* at 75° | $\lambda_{edge}$ at 0° |
|---|---|---|---|---|---|---|
| | % | % | | nm | nm | nm |
| SC-Cavity | 92 | - | 4 | 21 | 15 | 530 |
| OPF1 | 93 | 98 | 6 | 4 | 19 | 525 |
| OPF2 | 93 | 98 | 4 | 19 | 16 | 531 |

The high angular stability offered by polariton filters enables their implementation as a mechanically flexible filter film[62,63], which is impractical for standard *Fabry-Pérot* filters as light is incident on the filter over a multitude of different angles when the filter is flexed, thus severely affecting its performance. To realize mechanical flexibility, we sandwich a polariton filter between two highly flexible, ultrathin quasi-substrates formed by Parylene-C and a protection and adhesion layer of $Al_2O_3$ that we obtain by an optimized chemical vapor deposition process[64] (Fig. 4**a**). The resulting filters have a total thickness of less than 15 µm and can be reversibly applied to many surfaces, including flat and curved optical elements. The angle-resolved transmission of flexible versions of the metallic polariton filter from Fig. 2**d** and the dielectric OPF2 design from Fig. 3**f** are shown in Fig. 4**b** and **c**, respectively. Like their counterparts on glass, they exhibit largely angle-independent spectral characteristics. Due to thin film interference in the thin quasi-substrates, their transmission spectra show additional high-order parabolic modes at longer wavelengths. While not an issue in applications like fluorescence microscopy, if needed, this effect could be reduced by tuning the thickness of the Parylene-C layers or moving to thicker substrates. We emphasize that even the complex design of OPF2, with 55 dielectric and organic layers, can readily be implemented in a flexible form factor, in part because the symmetric Parylene-C quasi-substrates place the filter in the neutral plane of the stack, thus reducing mechanical stress.

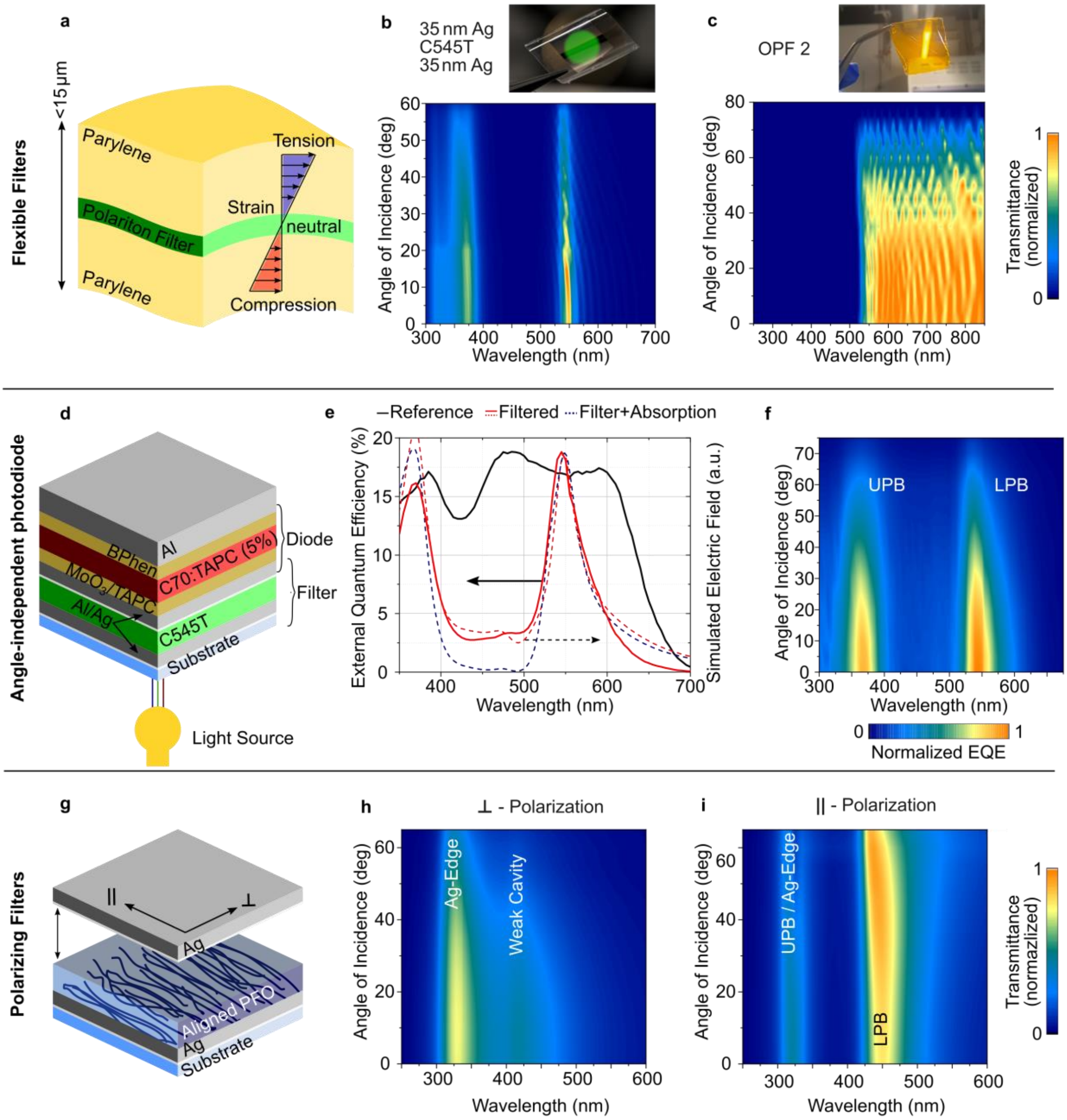

**Figure 4: Variations of polariton filters. a** Illustration of an ultrathin flexible filter design comprising two 6 µm thick parylene quasi-substrates and a C545T-based polariton filter. **b** Photograph and measured angle-resolved transmission of flexible metal-organic-metal polariton filter. **c** Photograph and measured angle-resolved transmission of a flexible variant of Optimized Filter 2 (OPF2). **d** Device structure of monolithically integrated photodiode and polariton-filter. **e** Experimental external quantum efficiency spectra (solid lines) of a reference photodiode (black) and the photodiode with integrated polariton filter (red) and corresponding simulated optical field amplitudes within the C70:TAPC heterojunction (dashed lines). The remaining response between 400 nm and 500 nm can be reduced by adding an absorbing C545T layer below the filter stack (simulation, dashed blue line). **f** Measured, angle-resolved external quantum efficiency of photodiode with integrated polariton filter, showing the angle-independent narrowband response at the positions of the UPB and LPB. **g** Illustration of polarizing polariton filter with strong coupling mediated by an aligned layer of PFO polymer. **h,i** Filter transmission for polarization perpendicular (**h**) and parallel (**i**) to the polymer chain orientation. Due to the polarization-dependent absorption of aligned PFO, strong coupling to the cavity mode only occurs for light polarized parallel to the polymer chains, leading to a polarization-dependent LPB and thus a highly polarization selective filter.

Next, we combine our polariton filters with broadband photodiodes. Instead of placing a filter at a macroscopic distance from the diode, we propose a monolithic design, where the diode is fabricated directly on top of the filter, with the light coming in through the substrate and the filter entering the diode over a range of angles (Fig. 4**d**). We use a C545T-based polariton filter and an organic photodiode (OPD) with 5 wt% TAPC diluted in $C_{70}$ as broadband absorber[65] (for diode characteristics see Supplementary Information, Fig. S17). The entire structure is produced in a single vacuum deposition run and has a total thickness of 320 nm. Figure 4**e** compares the EQE of a reference diode without filter to the diode comprising the polariton filter. While the reference shows a broadband spectral response, the EQE of the polariton filtered diode peaks at the positions of the LPB (450 nm) and UPB (380 nm) of the filter. Interestingly, due to the monolithic integration of filter and diode and the resulting optical coupling between them, the residual absorption of the polariton filter does not reduce the detection efficiency, and both diodes achieve comparable peak EQEs of ~18%. The coupling however also leads to a non-vanishing sensitivity between 400 nm and 500 nm, which is consistent with transfer matrix modelling of the standing electric field at the position of the $C_{70}$ layer (dashed red line in Fig. 4**e**). An additional, purely absorptive C545T can be integrated below the bottom mirror of the filter to avoid this effect (simulation, blue dashed line in Fig. 4**e**; a complete electric field simulation is presented in the Supplementary Information, Figure S18). Finally, to demonstrate the advantage of our design over integration of a conventional interference filter, we perform an angle-resolved measurement of the EQE, where we observe an angle-independent spectral response for the monolithic device as expected (Fig. 4f).

Lastly, we use molecular orientation in films of a semiconducting polymer to induce anisotropic strong coupling[38,66] and in turn create a polarization-dependent and angle-independent narrowband transmission filter (Fig. 4**g**). Poly(9,9'-dioctylfluorene) (PFO) is aligned by depositing it on top of a Ag mirror coated with an alignment layer of Sulfuric Dye 1 (SD1). Subsequently, the resulting film is heated above the liquid crystal transition temperature of PFO, followed by rapid quenching to room temperature to lock alignment[67]. Capping with a second Ag mirror completes the stack. As the transition dipole of PFO is mainly oriented along the polymer backbone, strong coupling is only induced for light polarized parallel to the alignment of the polymer chains. This effect leads to a drastically different transmission response for the two possible polarizations. For light polarized perpendicular to the chain orientation, the sample is in the weak coupling regime and shows only a weak cavity mode at a wavelength of around 420 nm (Fig. 4**h**). By contrast, for light polarized parallel to the chain orientation, a narrowband and nearly angle-independent LPB appears at around 450 nm (Fig. 4**i**).

## Discussion

In summary, we demonstrated exciton-polariton-based optical filters with stable angular response using metal-organic-metal and multilayer dielectric cavities carefully designed to exploit ultra-strong coupling. These filters match or outperform comparable conventional filters in many performance metrics and show a blueshift smaller than their linewidth even at extreme angles of incidence >80°(see Supplementary Information, Figure S16 and Table S2, for a comparison to existing filters). Using a series of organic materials, we realized filters operating throughout the visible spectral range and into the near infrared, with even longer wavelengths likely also accessible in the future with appropriate absorber materials. Furthermore, we showed that ultra-strong coupling in dielectric resonators allows the design of high extinction optical filters, with an OD of up to 6 demonstrated here.

The amorphous nature of organic materials further allowed fabrication of ultrathin filters that are mechanically flexible. In the future, such flexible filters might be mass-produced on carrier films to enable scalable and cost-efficient fabrication. These filters could then be applied to various non-planar optical components in a simple and reversible manner.

We also demonstrated monolithic integration of a polariton filter with a broadband organic photodiode to realize a thin-film photodetector (thickness <500 nm) with non-dispersive and narrowband spectral response. We expect that integration with different polariton filters will allow to tailor the response of integrated photodetectors to the requirements of a wide range of applications. Beyond the integration with organic photodetectors, polariton filters might also be deposited or laminated onto charge-coupled-device (CCD) and complementary metal-oxide semiconductor (CMOS) cameras, thus enabling the use of high numerical aperture optics in spectral detection and hyperspectral imaging[68].

Finally, we utilized anisotropic strong coupling to design polarizing polariton filters. Like for the non-polarizing filter designs above, using films of aligned materials with different absorption profiles[66] will allow to produce polarization-dependent polariton filters with tailored spectral characteristics.

The use of ultra-strong light-matter coupling to manipulate angular dispersion differs fundamentally from a system where an angle-dependent thin-film interference filter and angle-independent absorption filter are stacked on top of each other, in the sense that strong light-matter interaction redistributes the spectral bands defined by the interference rather than just adding absorption in a specific spectral region (see Supplementary Information Fig. S12 for a direct comparison of the two approaches). Among other advantages, this allows a reduction of the amount of absorptive material used and retains the design freedom offered by dielectric filters. In addition, polariton filters also show considerably improved stability against inhomogeneities and variations in film thickness, and due to their angle-independence, are more robust to misalignment than conventional dielectric filters.

The strong absorption of organic materials allows for the creation of ultrathin polariton filters (metal-organic-metal <200 nm, DBR <4 μm) and the breadth of organic materials facilitates the design of polariton filters with a wide range of spectral response. In addition, the concept is likely not limited to organic materials; strongly interacting perovskites, quantum dots and other low-dimensional systems, as well as inorganic structures can all be explored for different wavelength regimes and to add additional functionality.

In the future, our filter concept might be enhanced further, e.g. by introducing strongly absorbing thin films into more complex multilayer coatings with designs tailored to the needs of specific high-performance optics. Thin-film interference filters with minimal levels of angular dispersion have long been sought after by optical engineers and manufacturers of optical coatings, but thus far they can only be realized to a very limited degree by using expensive, high-index materials and high-precision deposition systems or by very extensive patterning. Our approach thus opens a new parameter space for optical coatings and the design of optical systems in general. We envision that angle-independent polariton filters will be of particular relevance to micro-optics, sensing[69], display applications[35], and biophotonics[70]. In all these areas, it is often impossible or impractical to work with collimated beams of light and, therefore, light is frequently incident on filter elements over an extended range of angles. Polariton-based optical filters will likely disrupt the design rules for such systems and allow for ground-up redesigns with improved performance as well as reduced size and complexity.

## Methods

*Sample fabrication:* Metal-organic-metal polariton filters were fabricated via thermal evaporation of organic and metal thin films at a base pressure of $1 \times 10^{-7}$ mbar (Angstrom

EvoVac) onto 1.1 mm-thick glass substrates. The materials used were Al and Ag as metallic reflectors (Kurt J. Lesker Co.), 2,2',7,7'-tetrakis(N,N'-di-p-methylphenylamino)-9,9'-spirobifluorene (Spiro-TTB), 4,4'-Bis(4-(9H-carbazol-9-yl)styryl)biphenyl (BSBCz), 2,3,6,7-tetrahydro-1,1,7,7,-tetramethyl-1H,5H,11H-10-(2-benzothiazolyl)quinolizino-[9,9a,1gh]-coumarin (C545T), 2,3,9,10,16,17-hexachlorinated boron subphthalocyanine chloride ($Cl_6$SubPc) and Boron sub-2,3-naphthalocyanine chloride (SubNc) as strong coupling layers and $MoO_3$, 1,1-Bis[(di-4-tolylamino)phenyl]cyclohexane (TAPC), Fullerene C70, 4,7-Diphenyl-1,10-phenanthroline (BPhen) for photodiode fabrication. All organic materials and $MoO_3$ were obtained from Lumtec in sublimed grade and used as received. For dielectric polariton filters, $Ta_2O_5$ and $SiO_2$ films were produced by radiofrequency magnetron sputtering from a $Ta_2O_5$ or $SiO_2$ target (Angstrom) at a base pressure of $1 \times 10^{-7}$ mbar, using 18 standard cubic centimeters per minute (sccm) Argon flow at 2 mTorr process pressure. For $Ta_2O_5$, an additional oxygen flow of 4 sccm was added to prevent the formation of suboxides. Film thicknesses were controlled in situ using calibrated quartz crystal microbalances (QCMs). Before fabrication, substrates were cleaned by ultrasonication in acetone, isopropyl alcohol and deionized water (10 min each), followed by $O_2$ plasma-ashing for 3 min. Flexible filters were prepared on cleaned glass carrier substrates, similar to the process described in Keum e*t al.*[64]. In brief, parylene-C (diX C, KISCO) and $Al_2O_3$ layers were deposited using a parylene coater (Parylene P6, Diener electronic) and an atomic layer deposition (ALD) reactor (Savannah S200, Veeco) which are connected to the evaporation chamber via a nitrogen filled glovebox. After processing, the devices were peeled off the carrier substrate to yield free-standing, flexible filters. Filters were used in air without further encapsulation. Photodiodes were encapsulated in a nitrogen atmosphere with a glass lid using UV-curable epoxy (Norland NOA68), preventing intermittent exposure to ambient air. The active area of these devices was 16.0 $mm^2$. For polarising filters, a layer of sulfuric dye 1 (SD1, Dai-Nippon Ink and Chemicals, Japan)[38] was spin-coated, annealed for 10 min at 150 °C, and exposed to 5 mW of polarized UV light for 10 min to photo-align the film. Films of poly(9,9-dioctylfluorene) (PFO, Sumitomo Chemical Company) were spin-coated on top of the SD1 film in an inert environment, using 8 mg $ml^{-1}$ PFO in toluene solution. The sample was placed on a precision hotplate in an inert environment and the temperature was raised from 25 to 160 °C at a rate of approximately 30 °C $min^{-1}$. The upper temperature was then held for 10 min, followed by rapid quenching to room temperature by placing the sample on a metallic surface.

*Computational design of multilayer stacks:* Positioning of the organic layers inside the multilayer stack was first optimised using a particle swarm optimization algorithm[59] combined with the needle method[71]. This was followed by a further optimization using the Levenberg-Marquardt algorithm implemented in OpenFilters[61]. The number and thickness of layers was restricted to ensure an experimentally viable design. When fabricating the optimized designs, a thickness uncertainty of ~5% is introduced due to fabrication imperfections related to QCM tooling and the complexity of the designs. Industrial optical coatings routinely achieve better thickness control than what we had access to for this work.

*Device characterization:* Optical constants of all used materials and angle-resolved transmission measurements of the filters were obtained with a J.A. Woollam M2000 varying angle spectroscopic ellipsometer. Measurements at high optical density were performed using a two-beam spectrophotometer (Lambda 1050+, Perkin-Elmer). For verificiation, measurements were repeated three times; there were no observable differences between measurements within the same sample. All transmission values shown represent external transmission, i.e. the total fraction of light making it through the entire filter, including its substrate. Photodiodes were evaluated using a xenon-lamp, combined with a monochromator (Oriel Cornerstone 130 1/8 m), a chopper and a lock-in amplifier (SR830 DSP, Stanford

Research Systems). The excitation intensity was monitored by a calibrated silicon reference diode. Device structures and electric field distributions of the photodiodes were simulated using a transfer matrix model. Polariton branches were calculated using a coupled oscillator model[57]. Photographs were taken using a digital single-lens reflex camera (Nikon D7100) with a macro lens (Sigma 105 mm F2.8 EX DG OS HSM).

## Supplementary Information

Supplementary Information (Supplementary Discussion 1-3, Tables S1, S2 and Figures S1-S18) is available, including additional filter measurements, transfer matrix simulations of thickness variations, performance analysis, literature overview on angle-stable filters, additional j-V characteristics and electric field simulations of photodiodes.

## Acknowledgements

We are grateful to Prof. Klaus Meerholz for providing access to the variable angle spectroscopic ellipsometry setup and to Prof. Donal Bradley and the Sumitomo Chemical Company for provision of PFO. This research was financially supported by the Alexander von Humboldt Foundation (Humboldt Professorship to M.C.G.) and the European Research Council under the European Union's Horizon Europe Framework Programme/ERC Advanced Grant agreement no. 101097878 (HyAngle) and the ERC grant agreement no. 864625 (ConTROL). A.M. acknowledges funding through an individual fellowship of the Deutsche Forschungsgemeinschaft (no. 404587082), from the European Union's Horizon 2020 research and innovation programme under the Marie Skłodowska-Curie grant agreement no. 101023743 (PolDev) and from the Bundesministerium für Bildung und Forschung (BMBF) within a GO-BIO initial project no. 16LW0454 (FluoPolar). B.S. acknowledges funding from the Research Foundation – Flanders (FWO) through a Senior Postdoctoral Fellowship no. 12AOC24N (SOFIA). F.L.R. acknowledges funding from the Alexander von Humboldt Foundation through a Humboldt Fellowship.

## Author Contributions

A.M. and M.C.G. designed the study. A.M. and S.H. fabricated all devices. A.M. measured and analysed ellipsometry and angle-resolved transmission data. A.M., B.S., and K.V. designed photodetector devices, which were measured by B.S.. A.M. and F.L.R. designed and fabricated polarizing filters. All authors evaluated and discussed the results. A.M. and M.C.G. wrote the manuscript with input from all authors.

## Data availability

The data that support the findings of this study are openly available via the St Andrews Research Portal.

**Competing interests**

A.M. and M.C.G. have applied for IP on polariton-based filters.

# Supplementary Information:

# Breaking the angular dispersion limit in thin film optics by ultra-strong light-matter coupling

Andreas Mischok[1,2,*], Bernhard Siegmund[3], Florian Le Roux[1], Sabina Hillebrandt[1,2], Koen Vandewal[3], Malte C. Gather[1,2,*]

*[1]Humboldt Centre for Nano- and Biophotonics, Department of Chemistry, University of Cologne, Greinstr. 4-6, 50939 Köln, Germany*

*[2]School of Physics and Astronomy, University of St. Andrews, North Haugh, St. Andrews KY16 9SS, United Kingdom*

*[3]Institute for Materials Research (IMO-IMOMEC), Hasselt University, Wetenschapspark 1, Diepenbeek, 3590 Belgium*

**contact: andreas.mischok@uni-koeln.de, malte.gather@uni-koeln.de*

**Contents**

**Supplementary Discussion 1**

In the most basic case, a bandpass filter comprise a thin film sandwiched by two reflective surfaces, such as thin metallic mirrors, where the thin film thickness and mirror reflectivity determine the resonance wavelength and finesse of the filter, respectively. In addition, for thin metal mirrors the reflected phase $\phi$ of light is a function of metal thickness and leads to a significant deviation from the basic resonance condition for such cavities; this effect can be described by an effective cavity thickness[1]: $d_{\mathrm{eff}} = d + \lambda_{\mathrm{res}}/2\pi n \, * (|\phi_{\mathrm{bot}}| + |\phi_{\mathrm{top}}|)$. This is taken into account for the device design of the devices shown in Fig. 2 of the main text.

Correct tuning of the strong coupling parameters is crucial to achieve the best possible performance. As discussed in the main manuscript, ideally polariton filters operate at a large coupling strength $\hbar\Omega_R/2$ (ideally in the ultra-strong coupling regime) and at a moderate positive detuning. A large value of $\hbar\Omega_R/2$ helps in two ways. First, a large separation of upper and lower polariton branches is desirable to split up the two resonances. In a filter application, a single transmissive resonance is often required. When the separation between UPB and LPB is large, it is easy to filter out the unwanted UPB transmission (if necessary) or utilise an additional absorber to reduce it (compare Fig. S2, where the UPB from a C545T cavity is absorbed by an additional SpiroTTB layer outside the cavity). Second, a large coupling strength will lead to a larger separation of the LPB from the bare exciton resonance. This avoids parasitic absorption of uncoupled excitons at the LPB resonance and therefore maintains high transmission. Finally, the high coupling strength tends to allow for a flatter angular dispersion, especially if the transmission band is to occur away from the exciton resonance.

Similarly, the positive detuning is useful to keep a largely angle-independent dispersion. For a large negative detuning, the LPB is highly photonic at lower angles and thus retains the strong angular dispersion of the cavity mode. By using a positive detuning, we aim to make the dispersion more excitonic and thus flatter. However, at the same time, the positive detuning should not be made too large as this will reduce the transmission of the LPB. Through optical simulations, we have empirically found a value of 100 meV to 200 meV to be ideal. We showcase the transition from low to high transmission further below.

The full filter stack comprises: 1 nm Al | 25 nm Ag | 80 nm C545T | 1 nm Al | 25 nm Ag, and is fabricated on a glass substrate via thermal evaporation. In comparison, the architecture of the conventional metal-dielectric-metal (MDM) filter is 1 nm Al | 25 nm Ag | 140 nm $SiO_2$ | 1 nm Al | 25 nm Ag and is fabricated by a combination of thermal evaporation (for metals) and magnetron sputtering (for $SiO_2$). In order to maximize resonator quality and transmission, the thin silver mirrors are supported by a seed or wetting layer of aluminium, which is known to support growth of smooth silver films with increased optical quality[2]. Resonator quality can be further tuned by adjusting the silver thickness (compare Fig. S1).

Figure S3 shows the full spectra of the polariton filters shown in Fig. 2h of the main article. In addition, we show simulated transmission spectra for a corresponding set of MDM filters with a $SiO_2$ spacer, tuned to have transmission peaks for normal incidence at similar wavelengths, in Figure S4. For these, the strong angular shift and resulting overlap of the transmission spectra severely limits the usable angular range.

A further advantage of polariton filters over conventional interference filters is their higher resistance to thickness variations which might be introduced by production errors or local inhomogeneities. Figure S5 shows the simulated transmission spectra of an MDM filter and a C545T-based polariton filter for a variation in core-layer thickness of ±5%. For this thickness variation, the polariton filter exhibits a 3-fold lower spectral shift than the MDM filter. In general, polariton filters are tunable by a variation in thickness, however at the cost of angular stability for red-shifted wavelengths and peak transmission for blue-shifted wavelengths. Figure S6 demonstrates this trade-off in the measured angle-resolved spectra for a set of BSBCz-based polariton filters with cavity thicknesses between 40 nm and 70 nm. In practice, thickness tuning alone would not be used over such a wide wavelength range; instead the chemical versatility of organic materials allows to choose well-matched absorber materials for the desired spectral response.

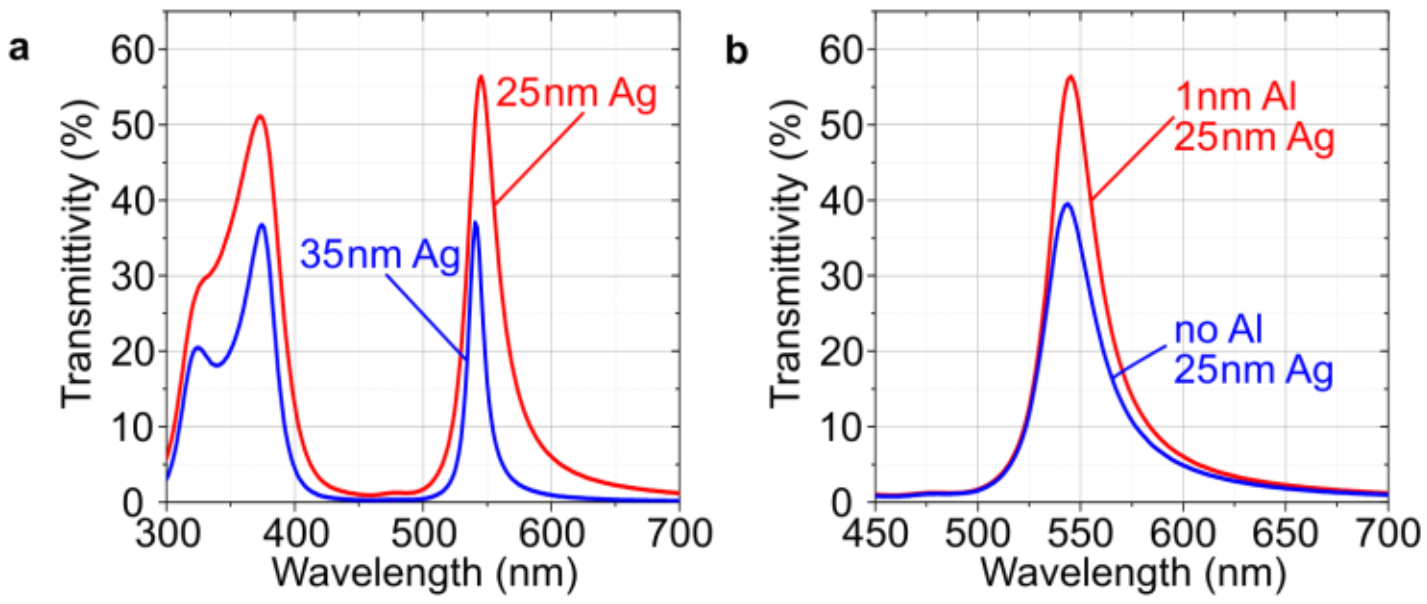

**Figure S1: Polariton filter transmission for different Ag mirrors**. **a** Transmission of a C545T-based polariton filter with Ag mirror of two different thicknesses. Increasing the Ag thickness leads to a narrower linewidth but lower transmission of the lower polariton mode. **b** Transmission of a C545T-based polariton filter with 25 nm thick Ag mirrors, with and without a 1 nm Al seed layer below each Ag layer. The Al seed leads to improved growth of Ag and thus a higher transmission (56% with vs 40% without Al in this example) and a sharper linewidth (27 nm FWHM with and 29 nm FWHM without Al), despite the poor optical properties of the Al layer itself.

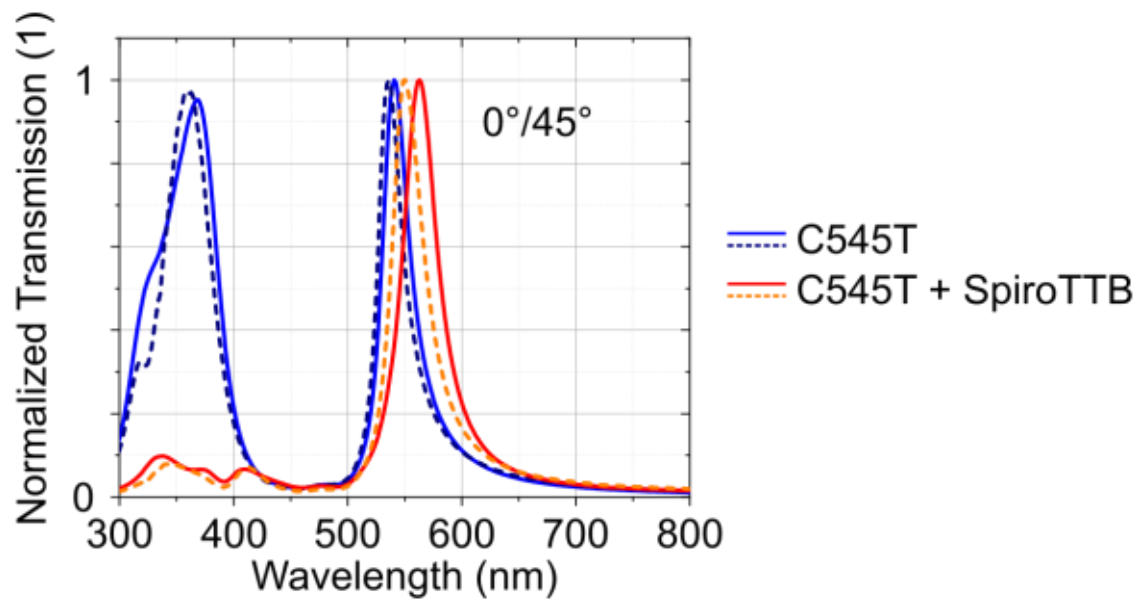

**Figure S2: Combination of a polariton-based filter with a Spiro-TTB absorption filter.** Normalized transmission of polariton filters with the structure 1nm Al | 25nm Ag | 80nm C545T | 1nm Al | 25nm Ag (blue lines) or 1nm Al | 25nm Ag | 90nm C545T | 1nm Al | 25nm Ag | 150nm SpiroTTB (red lines) at 0° (solid lines) and 45° (dashed lines) angle of incidence. The upper polariton branch of the C545T-based polariton filter shows high transmission at 350 nm, but this can be significantly reduced by the additional absorptive layer of Spiro-TTB.

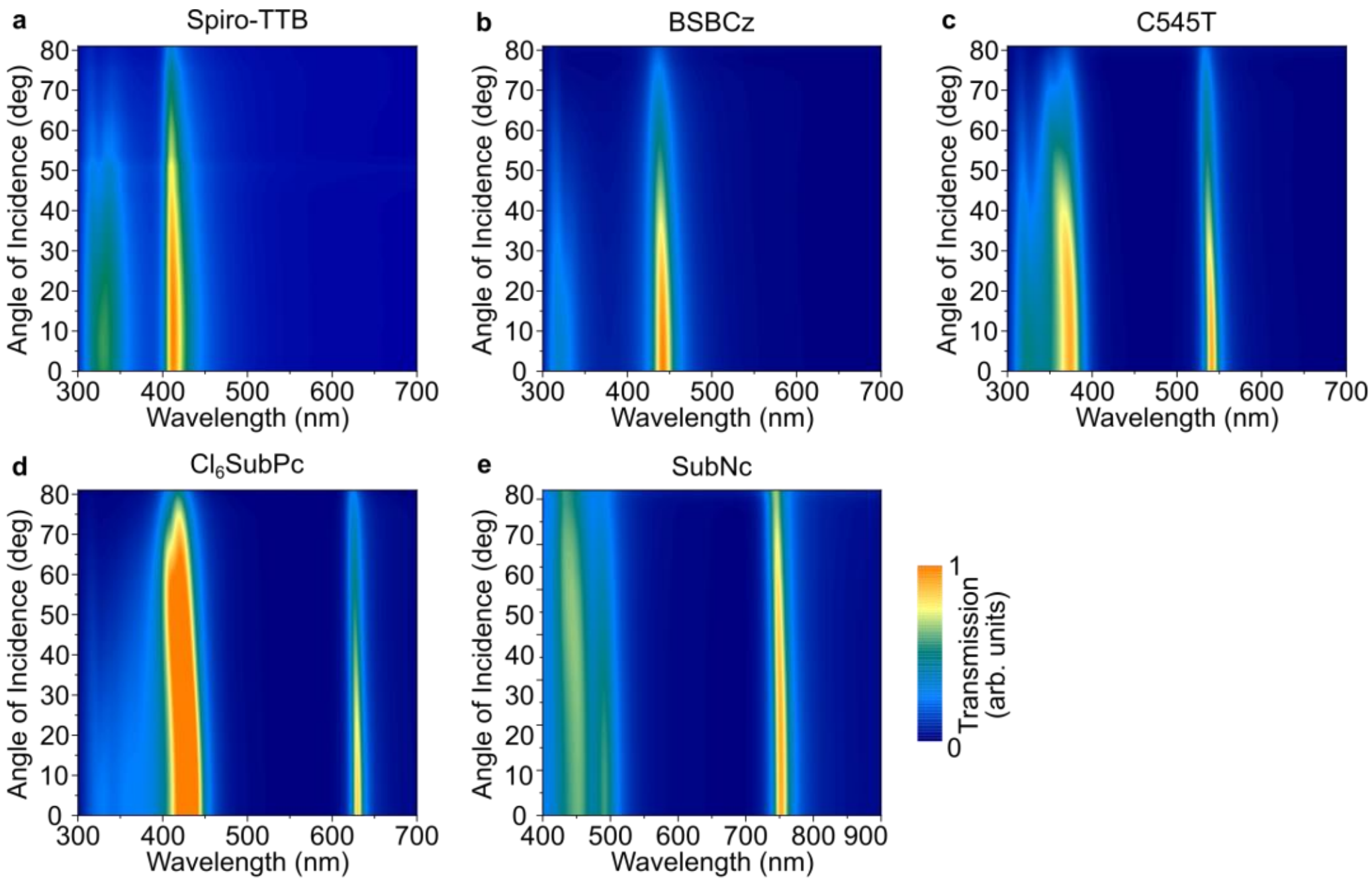

**Figure S3: Angle-resolved transmission for polariton filters with different materials.** Measured angle-resolved transmission of polariton-based filters with the structure: 1 nm Al | 35 nm Ag | SC layer | 1 nm Al | 35 nm Ag, with an optimized thickness of the SC layer tuned to the respective material and desired transmission wavelength: **a** 60 nm Spiro-TTB, **b** 50 nm BSBCz, **c** 90 nm C545T, **d** 80 nm $Cl_6SubPc$, and **e** 85 nm SubNc. By choosing an appropriate organic material with desired excitonic absorption, polariton filters with narrow and near angle-independent transmission bands across the visible spectral range can be fabricated. Spectra shown here correspond to data shown in Figure 3 of the main manuscript.

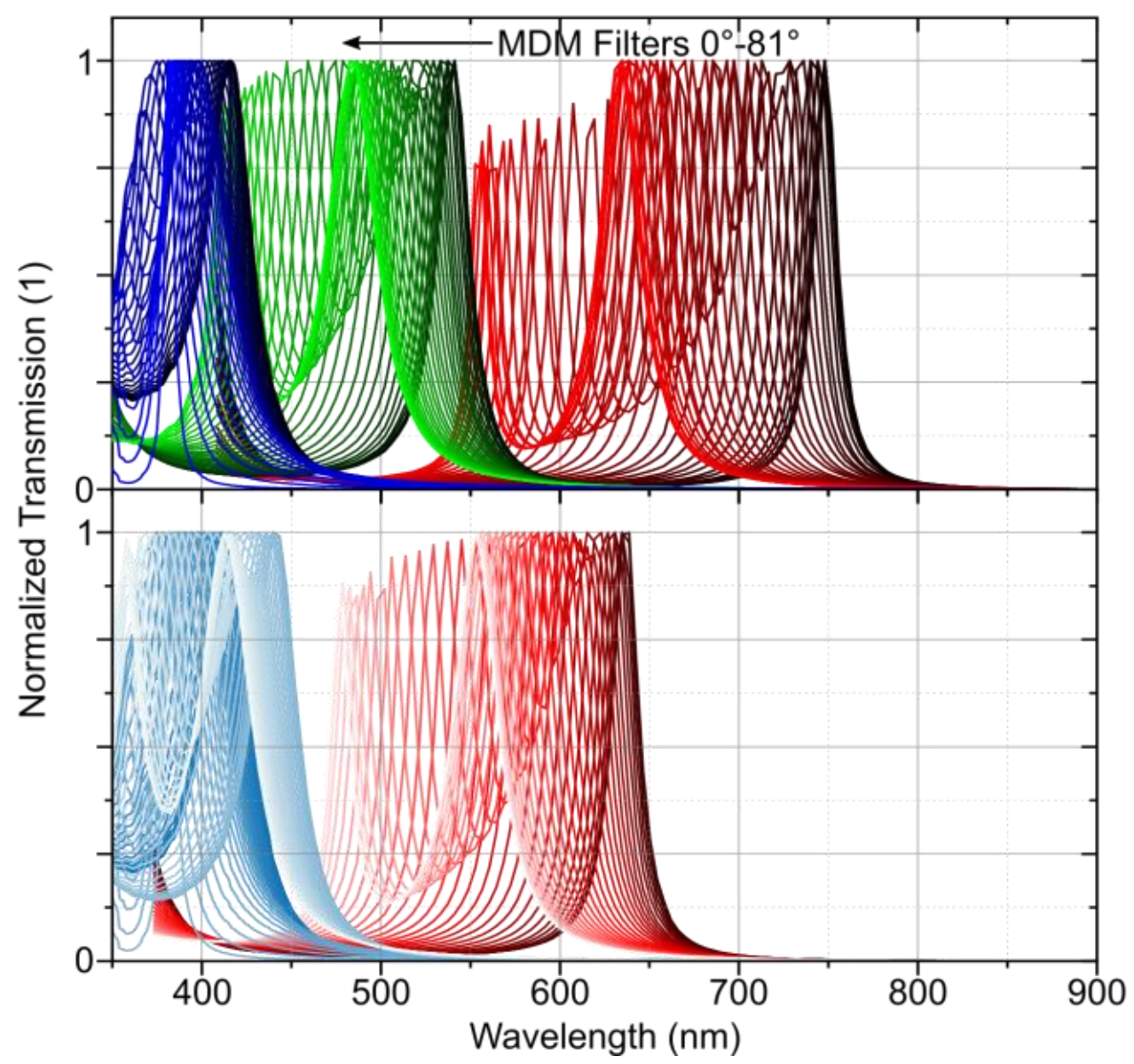

**Figure S4: Metal-dielectric-metal filters with transmission bands across the visible spectrum.** Simulated transmission spectra of MDM filters with the structure 1 nm Al | 35 nm Ag | xx nm $SiO_2$ | 1 nm Al | 35 nm Ag. The thickness of the $SiO_2$ layer is adjusted to between 95 nm and 210 nm, to yield peak transmission wavelengths at normal incidence that are similar to the polariton filters presented in Figure 3 of the main manuscript. The spectra of the MDM filters strongly shift with angle, and the spectra of the different filters show strong overlap at oblique angles, limiting their useful angular range.

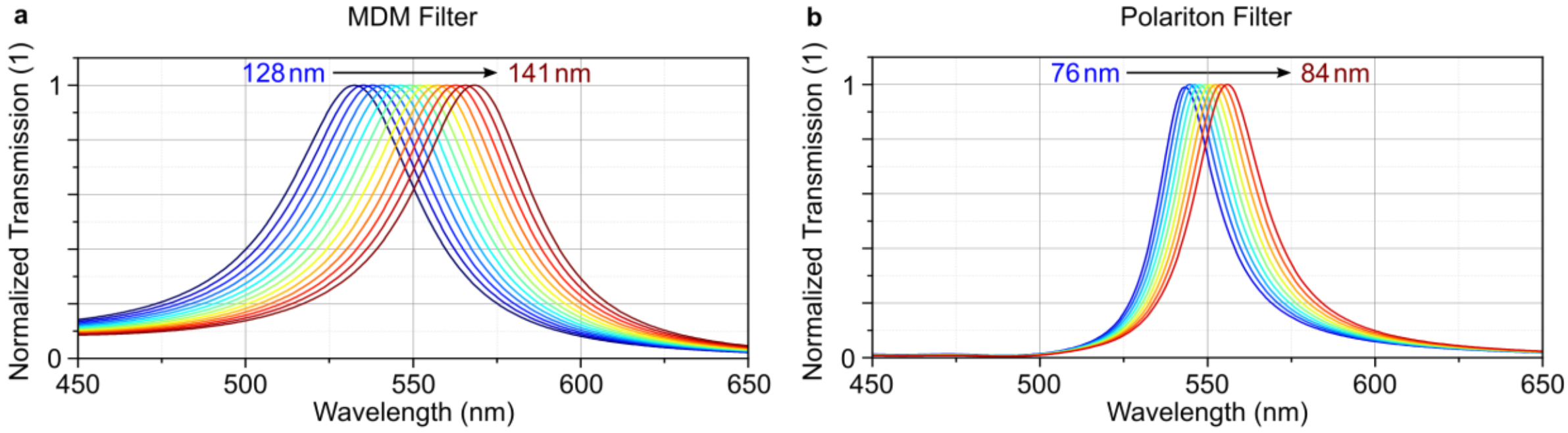

**Figure S5: Transmission of MDM and polariton filters for different core-layer thicknesses. a** Simulated transmission of an MDM filter with the structure 1 nm Al | 25 nm Ag | 135 nm $SiO_2$ | 1 nm Al | 25 nm Ag and with a thickness variation of the core layer of ±5% (arrow). This thickness variation leads to a spectral shift of 35.6 nm in the resulting bandpass filter. **b** Simulated transmission of a C545T-based polariton filter with the structure 1 nm Al | 25 nm Ag | 80 nm C545T | 1 nm Al | 25 nm Ag and with a thickness variation of the core layer of ±5% (arrow).The polariton filter shows a higher robustness to thickness variation with a spectral shift of only 12.5 nm.

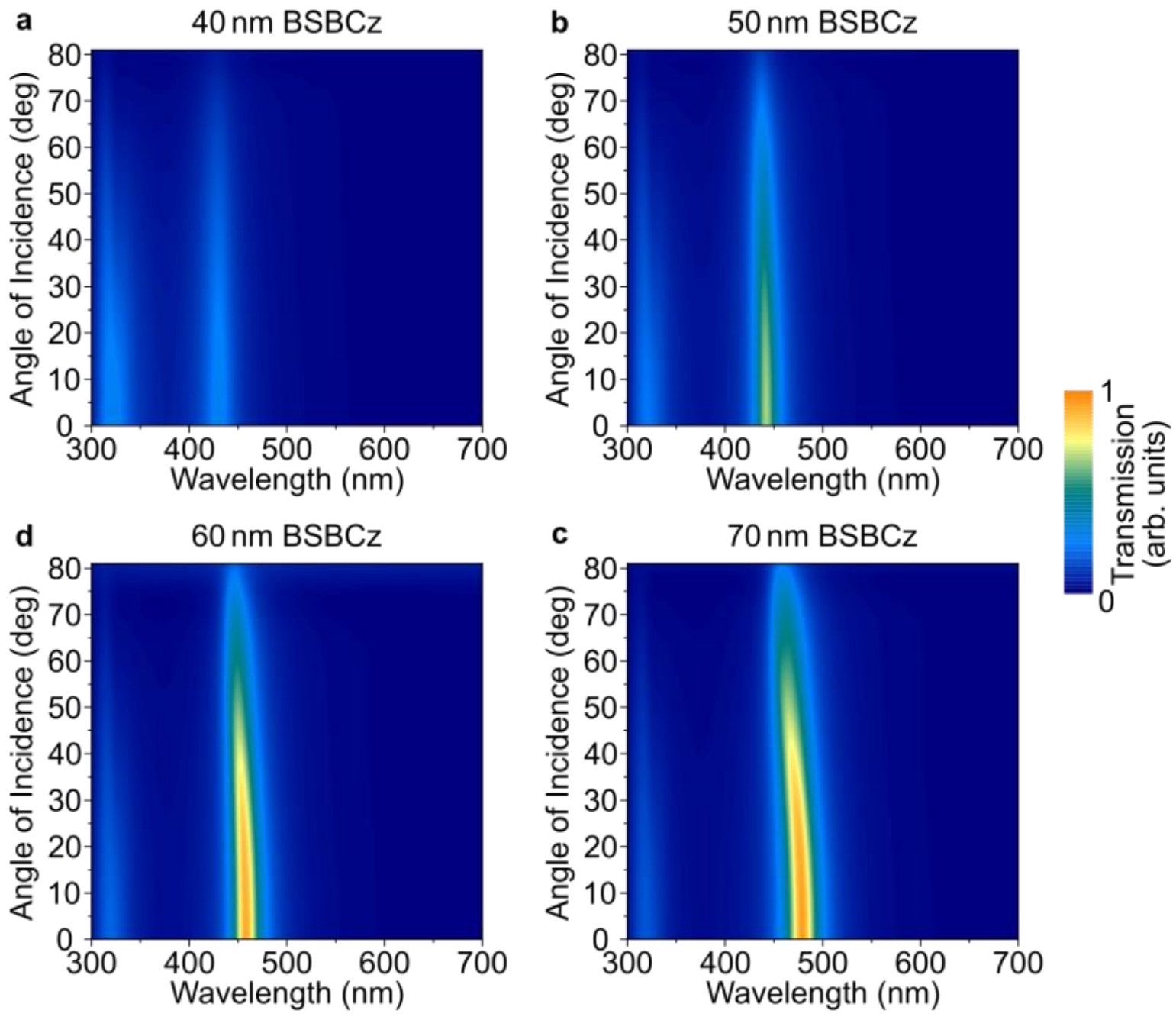

**Figure S6: Transmission of a BSBCz-based filter with different thicknesses.** Measured transmission spectra of BSBCz-based polariton filters with a BSBCz core layer thickness of **a** 40 nm, **b** 50 nm, **c** 60 nm, and **d** 70nm. A thickness variation of the organic core layer allows to tune the spectral position of the transmitted lower polariton branch, however at the cost of reduced peak transmission for blue-shifted resonances and an increased angular dependence for the most red-shifted resonances. In practice, a well-matched absorber material can be selected for the required spectral response, such that the thickness and associated detuning provide an ideal balance between peak transmission and angular stability.

**Table 1: Layer stacks for the optimized DBR-based polariton filters.**

| optimized Filter 1 (OPF1) | | | optimized Filter 2 (OPF2) | | |
|---|---|---|---|---|---|
| **Layer #** | **Material** | **Thickness (nm)** | **Layer #** | **Material** | **Thickness (nm)** |
| 1 | $Ta_2O_5$ | 44.5 | 1 | $Ta_2O_5$ | 36.6 |
| 2 | $SiO_2$ | 55.0 | 2 | $SiO_2$ | 63.5 |
| 3 | $Ta_2O_5$ | 57.8 | 3 | $Ta_2O_5$ | 52.2 |
| 4 | $SiO_2$ | 64.1 | 4 | $SiO_2$ | 81.2 |
| 5 | $Ta_2O_5$ | 64.1 | 5 | $Ta_2O_5$ | 52.9 |
| 6 | $SiO_2$ | 62.3 | 6 | $SiO_2$ | 76.3 |
| 7 | $Ta_2O_5$ | 59.7 | 7 | $Ta_2O_5$ | 52.2 |
| 8 | $SiO_2$ | 67.6 | 8 | $SiO_2$ | 77.3 |
| 9 | $Ta_2O_5$ | 53.0 | 9 | $Ta_2O_5$ | 34.8 |
| **10** | **C545T** | **100.0** | **10** | **Spiro:TTB** | **147.0** |
| 11 | $Ta_2O_5$ | 40.9 | 11 | $Ta_2O_5$ | 40.4 |
| 12 | $SiO_2$ | 69.7 | 12 | $SiO_2$ | 67.4 |
| 13 | $Ta_2O_5$ | 64.6 | 13 | $Ta_2O_5$ | 43.4 |
| 14 | $SiO_2$ | 58.7 | 14 | $SiO_2$ | 50.8 |
| 15 | $Ta_2O_5$ | 69.5 | 15 | $Ta_2O_5$ | 22.5 |
| 16 | $SiO_2$ | 59.2 | 16 | $SiO_2$ | 55.1 |
| 17 | $Ta_2O_5$ | 64.7 | 17 | $Ta_2O_5$ | 44.5 |
| 18 | $SiO_2$ | 65.0 | 18 | $SiO_2$ | 68.1 |
| 19 | $Ta_2O_5$ | 48.5 | 19 | $Ta_2O_5$ | 35.1 |
| **20** | **C545T** | **100.0** | **20** | **Spiro:TTB** | **152.5** |
| 21 | $Ta_2O_5$ | 55.0 | 21 | $Ta_2O_5$ | 37.7 |
| 22 | $SiO_2$ | 71.2 | 22 | $SiO_2$ | 75.2 |
| 23 | $Ta_2O_5$ | 62.9 | 23 | $Ta_2O_5$ | 48.3 |
| 24 | $SiO_2$ | 60.4 | 24 | $SiO_2$ | 65.6 |
| 25 | $Ta_2O_5$ | 69.0 | 25 | $Ta_2O_5$ | 36.0 |
| 26 | $SiO_2$ | 61.4 | 26 | $SiO_2$ | 22.6 |
| 27 | $Ta_2O_5$ | 62.3 | 27 | $Ta_2O_5$ | 41.7 |
| 28 | $SiO_2$ | 61.5 | 28 | $SiO_2$ | 64.7 |
| 29 | $Ta_2O_5$ | 47.5 | 29 | $Ta_2O_5$ | 54.7 |
| **30** | **C545T** | **100.0** | 30 | $SiO_2$ | 76.5 |
| 31 | $Ta_2O_5$ | 55.8 | 31 | $Ta_2O_5$ | 50.5 |
| 32 | $SiO_2$ | 71.1 | 32 | $SiO_2$ | 60.0 |
| 33 | $Ta_2O_5$ | 54.0 | 33 | $Ta_2O_5$ | 22.1 |
| 34 | $SiO_2$ | 63.2 | 34 | $SiO_2$ | 29.0 |
| 35 | $Ta_2O_5$ | 64.6 | 35 | $Ta_2O_5$ | 39.9 |
| 36 | $SiO_2$ | 72.0 | 36 | $SiO_2$ | 63.2 |
| 37 | $Ta_2O_5$ | 52.7 | 37 | $Ta_2O_5$ | 32.4 |
| 38 | $SiO_2$ | 62.2 | **38** | **C545T** | **142.6** |
| 39 | $Ta_2O_5$ | 67.9 | 39 | $Ta_2O_5$ | 37.8 |
| 40 | $SiO_2$ | 89.6 | 40 | $SiO_2$ | 70.6 |
| 41 | $Ta_2O_5$ | 13.5 | 41 | $Ta_2O_5$ | 48.3 |
| | | | 42 | $SiO_2$ | 68.1 |
| | | | 43 | $Ta_2O_5$ | 69.9 |
| | | | 44 | $SiO_2$ | 68.5 |
| | | | 45 | $Ta_2O_5$ | 47.1 |
| | | | 46 | $SiO_2$ | 70.3 |
| | | | 47 | $Ta_2O_5$ | 29.4 |
| | | | **48** | **C545T** | **147.8** |
| | | | 49 | $Ta_2O_5$ | 37.4 |
| | | | 50 | $SiO_2$ | 61.8 |
| | | | 51 | $Ta_2O_5$ | 46.5 |
| | | | 52 | $SiO_2$ | 78.1 |
| | | | 53 | $Ta_2O_5$ | 56.9 |
| | | | 54 | $SiO_2$ | 43.1 |
| | | | 55 | $Ta_2O_5$ | 57.6 |

## Supplementary Discussion 2

The introduction of organic materials into dielectric filters can lead to significantly enhanced angular performance. Figure S7 shows a transfer matrix model simulation that compares a 41-layer DBR to the 41-layer OPF1 from the main text. While the maximum optical density of both filters in the stopband is comparable, the strongly coupled filter shows a much improved angular stability, with a continuous stopband between 400 nm and 500 nm even at extreme angles of incidence. By contrast, the stopband of the conventional DBR shifts completely out of its original position for large angles. While the experimentally achieved performance of OPF1 is slightly decreased relative to the optical simulation (compare main text, Fig. 3**c**), the experimental behaviour is consistent with the simulated performance.

Figure S8 further elucidates the angular behaviour for the conventional DBR and for OPF1 and OPF2 using simuations of the electric field distribution inside the filters for light entering from the top side, first at 0° and in a second simulation at 60° angle of incidence. A gradual decrease in field intensity can be seen in the stopband for both the conventional DBR and the OPF1. However, while the stopband shifts drastically for the conventional DBR at an angle of incidence of 60°, the evolution of the electric field is relatively independent of angle in the strongly coupled OPFs.

Figure S9 shows the experimentally obtained performance of optimized Filter 2 on glass (Fig. S9**a**) and as a freestanding, flexible filter (Fig. S9**b**). The performance, in particular the OD, is not compromised by going to the flexible design.

Both metal and dielectric polariton filters show no mixing of polarization. Figure S10 shows the lack of cross-polarized sp and ps transmission at 0° and 45° incidence, comfirming no polarizing mixing or scrambling takes place in the polariton-based filters.

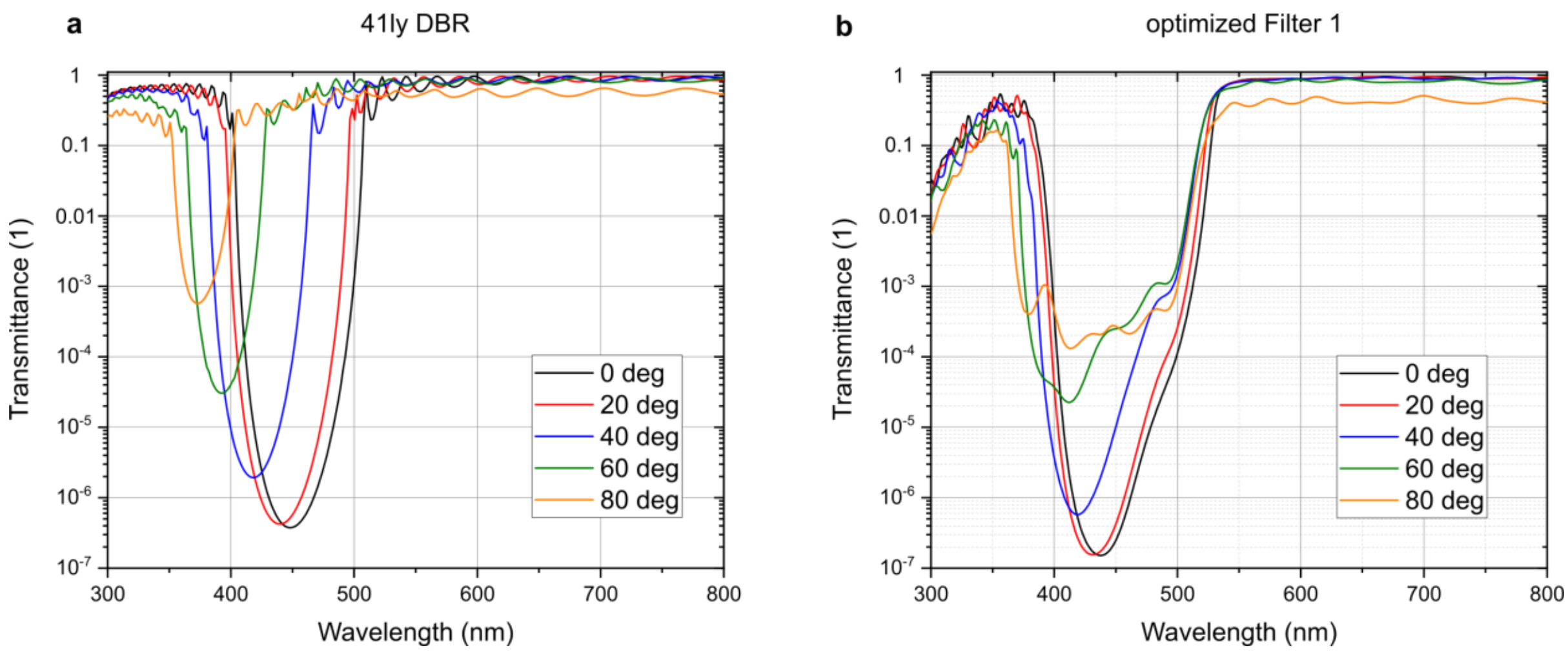

**Figure S7: Simulated transmission comparing DBR and Optimized Filter 1 (OPF1). a** Transmittance of a 41-layer conventional DBR at different angles of incidence. **b** Transmittance of OPF1 (also with 41 layers) at different angles of incidence. The performance at normal incidnece is comparable to the conventional DBR but the angular stability of the stopband is significantly increased.

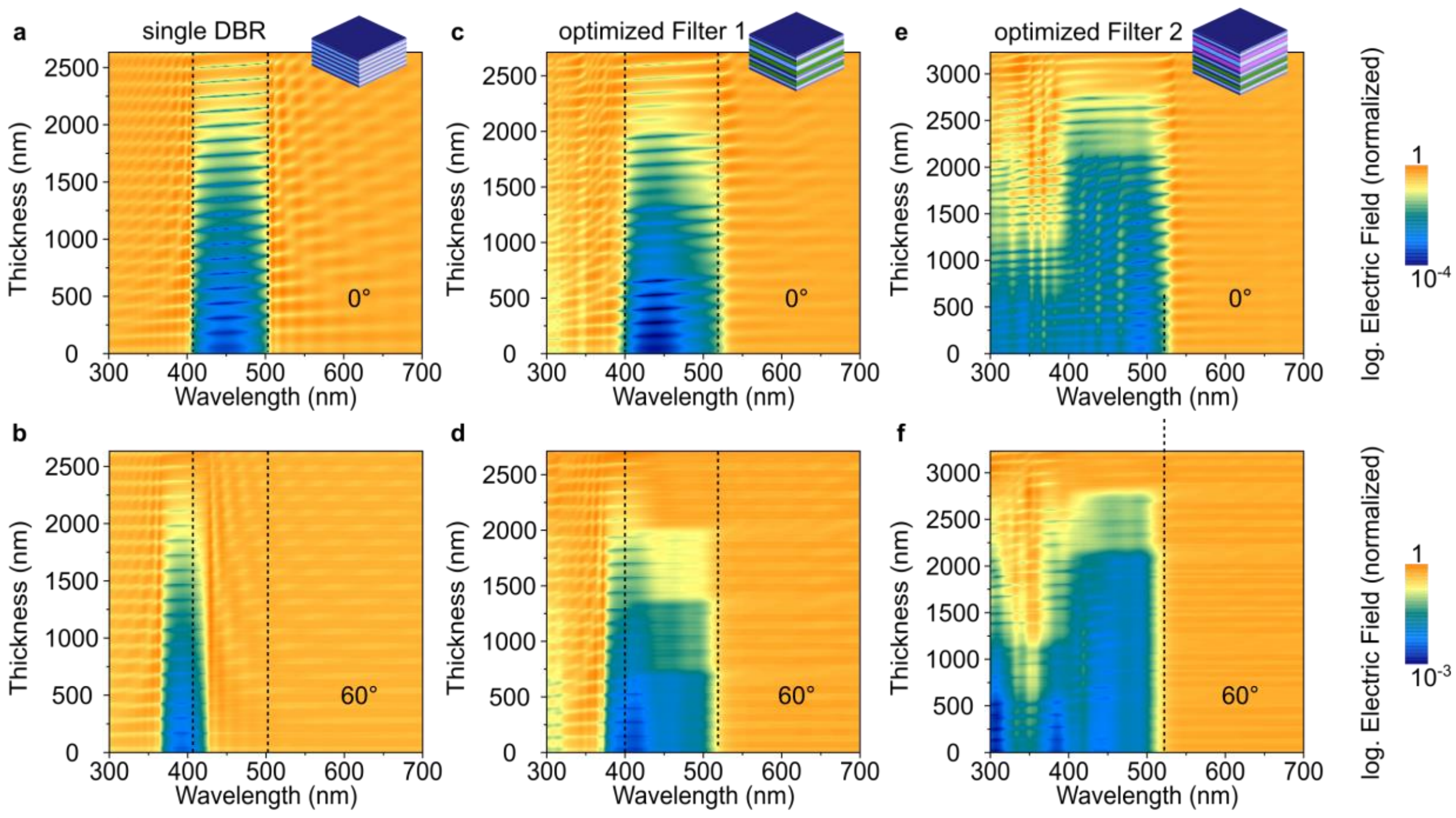

**Figure S8: Electric field simulation of DBR-based filters. a,b** Electric field in a 41-layer conventional DBR with light incident from the top at 0° (**a**) and 60° (**b**) angle of incidence. Dashed lines are a guide to the eye indicating the stopband position. At 60°, the stopband becomes much narrower and almost completely shifts outside of its initial position. **c,d** Electric field in OPF1 (also with 41 layers) with light incident from the top at 0° (**c**) and 60° (**d**) angle of incidence. Utilizing 3 organic cavities with C545T slightly broadens the stopband and renders it largely angle-independent. **e,f** Electric field in OPF2 (with 55 layers) with light incident from the top at 0° (**e**) and 60° (**f**) angle of incidence. The stopband can be broadened further by a stacked DBR design with 2x2 organic cavities of Spiro:TTB and C545T, creating a broadband, angle-independent longpass filter.

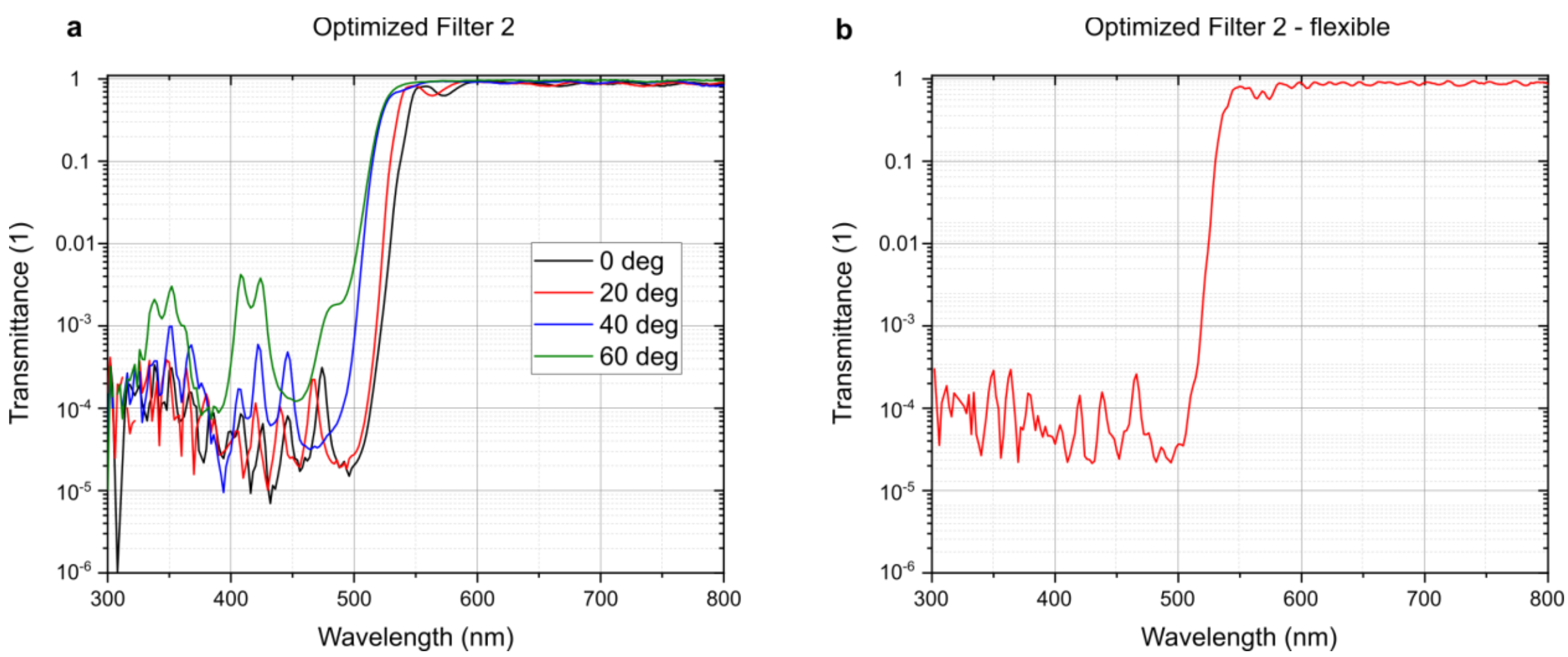

**Figure S9: Transmission of Optimized Filter 2 (OPF2). a** Measured transmittance of OPF2 on logarithmic scale for different angles of incidence. **b** Transmittance of OPF2 in a flexible, substrate-less design using Parylene-C carrier layers.

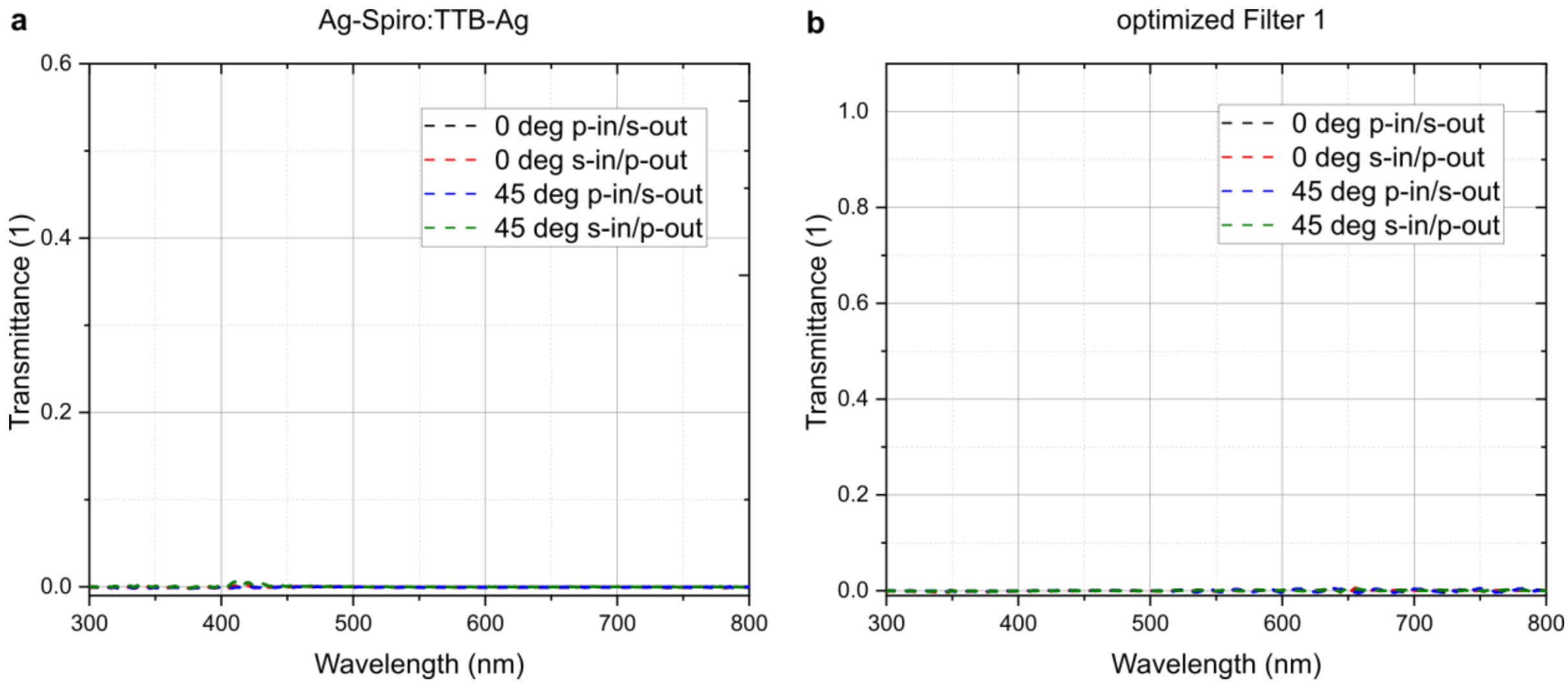

**Figure S10: Polarization mixing of polariton-based filters. a,b** Measured transmission of light with linear cross-polarization (s,p) of incident (in) and detected (out) light when passing through a metal-organic-metal SC filter (**a**) and the DBR-based Optimized Filter 1 (**b**) at 0° and 45° angle of incidence. The transmitted light shows no signs of polarization mixing or scrambling of polarization when passing through the polariton-based filters but instead maintains its linear polarization.

**Supplementary Discussion 3 – Performance limits and comparison of polariton filters**

To utilize the advantageous properties of polariton filters, their performance needs to be competitive when compared to the state-of-the-art in conventional optical coatings. In order to fully assess the performance achievable with polariton filters, we performed transfer matrix simulations in order to ensure that the discussion of fundamental limits not to be impacted by losses related to fabrication imperfections – such as enhanced plasmonic absorption by non-ideal silver film quality. Figure S11 showcases metal-metal polariton filters with optimized peak transmission, using ideal thin silver films[3] of 25 nm thickness. For such filters the peak transmission can easily reach values above 70% and up to 84%, depending on material, even without further enhancement by the use of anti-reflection layers, which are commonly used in conventional optical coatings.

One might assume that the introduction of organic materials to achieve strong coupling introduces additional absorption losses. However, we find that transfer matrix simulations of conventional MDM filters that use the same silver mirrors as above reach a similar performance (76% peak transmission for a $SiO_2$ core, 82% for a $Ta_2O_5$ core). Further analysis indeed shows that the remaining losses in our metal-metal polariton filters originate mainly from residual absorption in the metal layers and from reflection at the outer air-silver interface. Absorption in the organic materials does not represent a significant loss pathway (absorption at peak <3%). When the filter is properly tuned, parasitic absorption from the lower polariton can be almost fully avoided due to its redshift compared to the bare exciton.

Furthermore, the transition from photonic to polaritonic dispersion not only enables a mode with excellent angle stability, it also leads to higher peak transmission at larger angles. Fig. S11 **b** shows the maximum transmission of a SubPc-based polariton filter versus angle compared to a conventional MDM filter with a $SiO_2$ core. The polariton filter maintains a transmission above 75% for angles of incidence up to 50° and above 60% transmission for all angles, while the conventional filter drops to 50% at 50° and 45% above 60°. We believe this behaviour is due to the high refractive index that organic materials show on the red side of their excitonic absorption band because of the *Kramers-Kronig* relation. In this context, we would also like to point out that while the nominal transmission inevitably decreases with increasing angle of incidence for all filters, this effect is more than compensated by the much larger solid angle available at higher angles of incidence. For example, for a beam of light homogeneously covering +/-60° angle of incidence, 73% of the total light intensity is in the 30...60° cone. Therefore, a small drop in transmission at large angles does not significantly diminish the usefulness of having access to these larger angles.

The polariton filters presented here fundamentally differ from a simple combination of an MDM filter and an absorptive filter. For the latter, the blue-shifted transmission at higher angles would simply be absorbed, resulting in a sharp drop in transmission without a remaining detectable mode. Instead, in polariton filters, light is redirected into the polariton branches. In this way, the LPB retains a high and spectrally stable transmission over a much larger angle range, while transmission in the blocking region between LPB and UPB is still strongly suppressed. To demonstrate the difference between a polariton filter and a combination of an MDM structure and an organic absorbing material placed outside MDM cavity, Figure S12 shows the angle-resolved transmission of a SubPc-based polariton filter, a conventional MDM filter with $SiO_2$ as the dielectric, a bare SubPc layer used as a purely absorptive filter, as well as the combination of MDM and organic absorptive filter located outside the MDM cavity; for the latter, we distinguish the case where the organic material is outside but in direct contact with the MDM cavity and the case where the organic materials and the MDM cavity are on opposite sides fo the substrate. This comparison demonstrates that only the polariton filter shows a strongly improved angular performance, allowing for transmission of light at a stable wavelength even at high angles, while all other filter variants show a strong shift in wavelength with angle, a broad and unspecific transmission spectrum, or a dramatic loss in absolute transmission at larger angles.

While there is an optimal position of the polariton mode to achieve the best possible filter performance for any given material, there is nevertheless significant room to adjust the spectral position of the transmitted mode, while maintaining high transmission and low angular dispersion. Figure S13 explores the tunability of an optimized SubPc-based polariton filter, changing the core layer thickness from 50 nm to 100 nm. In turn, the peak transmitted wavelength shifts by ~80 nm. While at low thickness, i.e. for severely blue-shifted cavities, the transmission is drastically reduced, we observe a stable high transmission for core layer thicknesses >70 nm, representing a spectral tuning range of >50 nm over which we maintain excellent angular stability. The increase in transmission for increasing cavity thickness is directly related to reducing the parasitic absorption from the organic layer through uncoupled excitons. Above 70 nm, this parasitic absorption vanishes completely, and the remaining losses are due to (i) parasitic absorption in the metal layer and (ii) Fresnel-reflection at the substrate-air interface. Neither of these can be further reduced by a change in cavity thickness and neither is related to the presence of the organic material inside the cavity. This thus represents the best possible transmission in such a design.

Figure S14 showcases the polarization-resolved transmission of the SubPc-based filter compared to a conventional MDM filter. While the conventional filter shows a strong angle-dependence and polarization-splitting at large angles, the polariton filter shows a stable spectrum for both polarizations. Due to the different interaction of the two polarizations with the mirror layers, both filters show a higher transmission in p-polarization compared to s-polarization at high angles.

Finally, dielectric thin film stacks offer the best optical performance for both conventional and polariton-based filters. The Optimized Polariton Filter 1 (OPF1) shows a measured peak transmission of 93%, with the remaining loss largely due to Fresnel reflections at the back substrate-air interface. Figure S15 **a** shows that this reflection can be efficiently reduced by using a single or multilayer backside anti-reflection coating (ARC). Even the addition of just a single low-index film of $MgF_2$ increases the peak transmission to 96%, while a 9-layer ARC further enhances transmission to 98%. By simultaneously optimizing the ARC and the front-side filter, we can further enhance the performance as shown in Fig. S15 **b**, leading to an improved design that exhibits >98% transmission in its passband for angles up to 40° and >93% for angles up to 60°, without any major spectral shift of the edge onset.

We are therefore confident that such designs will compare very well against any conventional high-performance multilayer stack and that they outperform current approaches toward angular stability. To quantify this further, Figure S16 and Table S2 compare the performance of our polariton filters to solutions discussed in the current scientific literature. We find that polariton filters reach or exceed the performance of current solutions while keeping design complexity and fabrication effort low, especially when compared to approaches requiring nano-scale patterning, such as plasmonics and metasurfaces. The cost of the common organic materials that we used to induce strong coupling is also much lower than that of high-index materials, which are the basis of the best performing current alternatives used in research and industry.

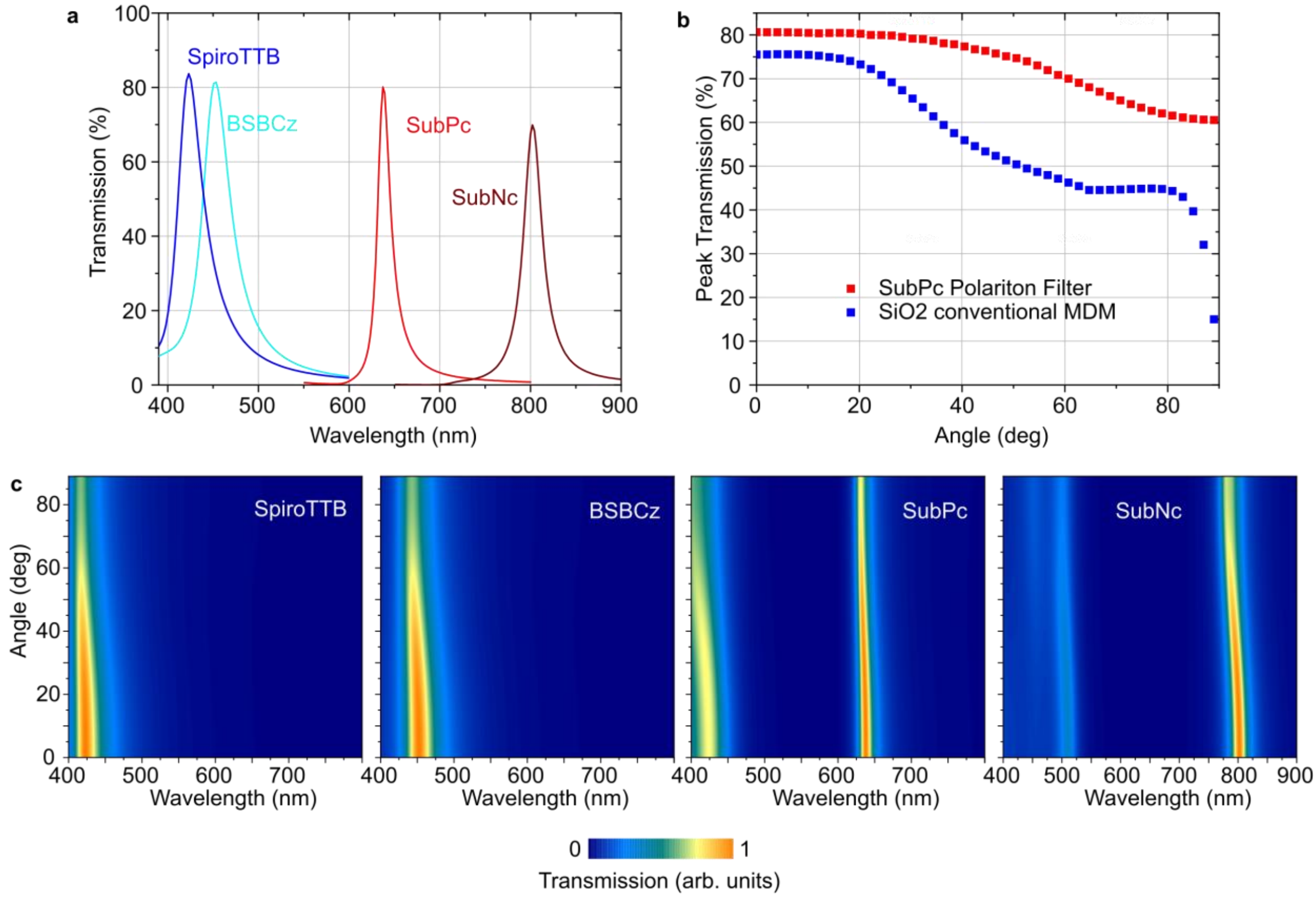

**Figure S11 Ag-Ag polariton filters with optimized peak transmission. a** Simulated transmission spectra for a series of metal-organic-metal polariton filters tuned to different wavelengths and optimized for peak transmission. Data obtained by transfer matrix simulations, assuming silver films of 25 nm thickness and with ideal optical properties[3]. **b** Simulated peak transmission vs angle for a SubPc-based polariton filter and a conventional $SiO_2$-based MDM filter tuned to the same resonance wavelength at 0° angle of incidence. The polariton filter maintains a higher peak transmission at large angles even when taking the spectral red-shift of the conventional cavity into account. **c** Simulated angle-resolved transmission spectra for the filters shown in **a** and **b**.

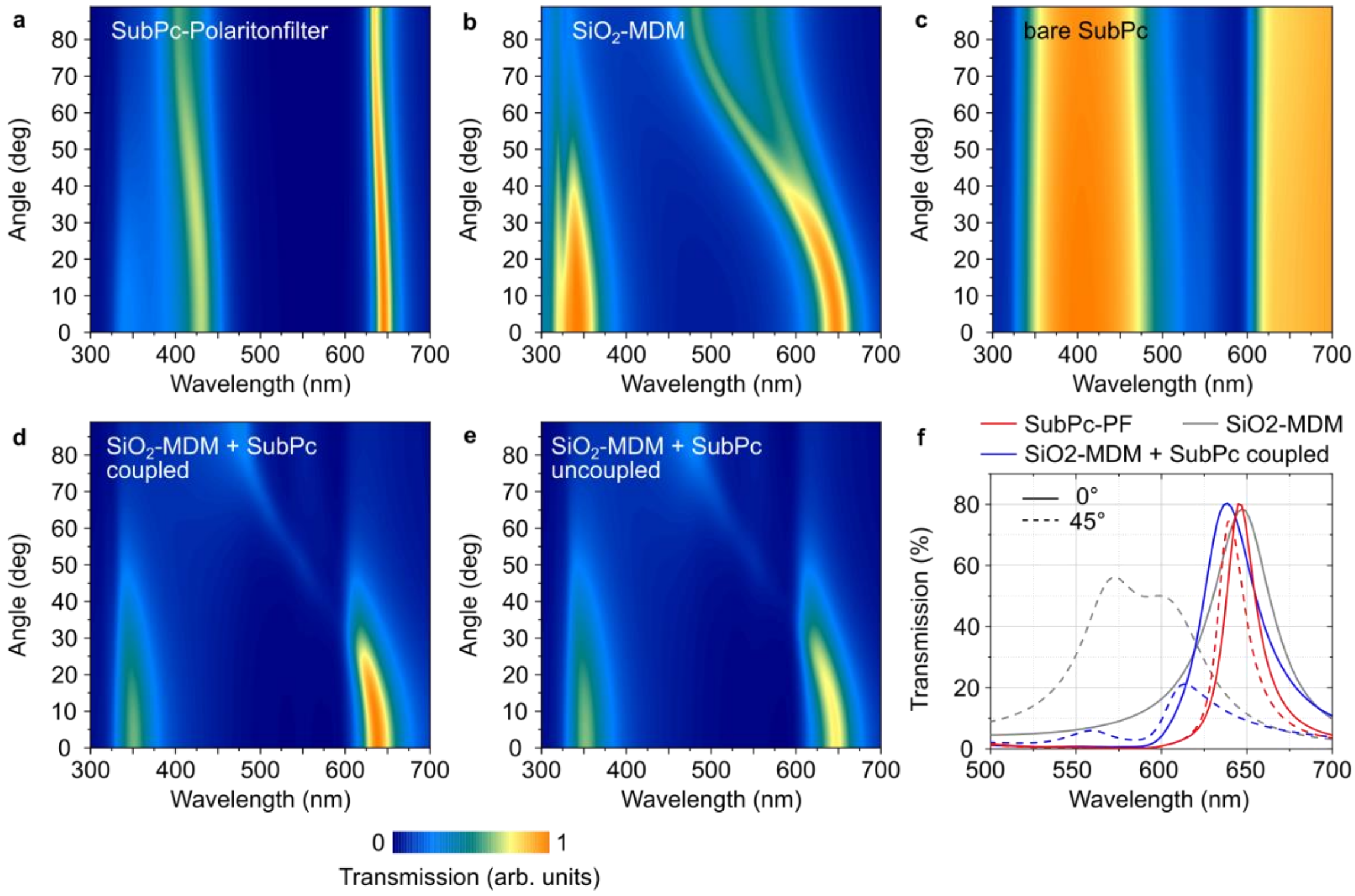

**Figure S12 Comparing polariton filters against MDM filters, absorptive filters and combined MDM / absorptive filters. a**-**e** Angle-resolved transmission of **a** 25 nm Ag | SubPc | 25nm Ag polariton filter, **b** 25 nm Ag | $SiO_2$ | 25nm Ag MDM-filter, **c** 80 nm SubPc absorptive Filter, **d** 80nm SubPc | 25 nm Ag | $SiO_2$ | 25nm Ag combined absorptive MDM filter with both elements directly next to each other allowing for some degree of optical coupling between the layers, **e** 80nm SubPc | Substrate | 25 nm Ag | SubPc | 25nm Ag combined absorptive/MDM filter without optical coupling between the two elements. **f** Direct comparison of SubPc polariton filter (red), $SiO_2$-MDM (grey) and coupled $SiO_2$-MDM + SubPc (blue) at 0° (solid lines) and 45° (dashed lines) angle of incidence. The polariton filter acts fundamentally different to a combination of a conventional dielectric and absorptive filters, as the transmission is redirected towards the polariton branches instead of just being passively absorbed, resulting in a high and spectrally stable transmission at all angles.

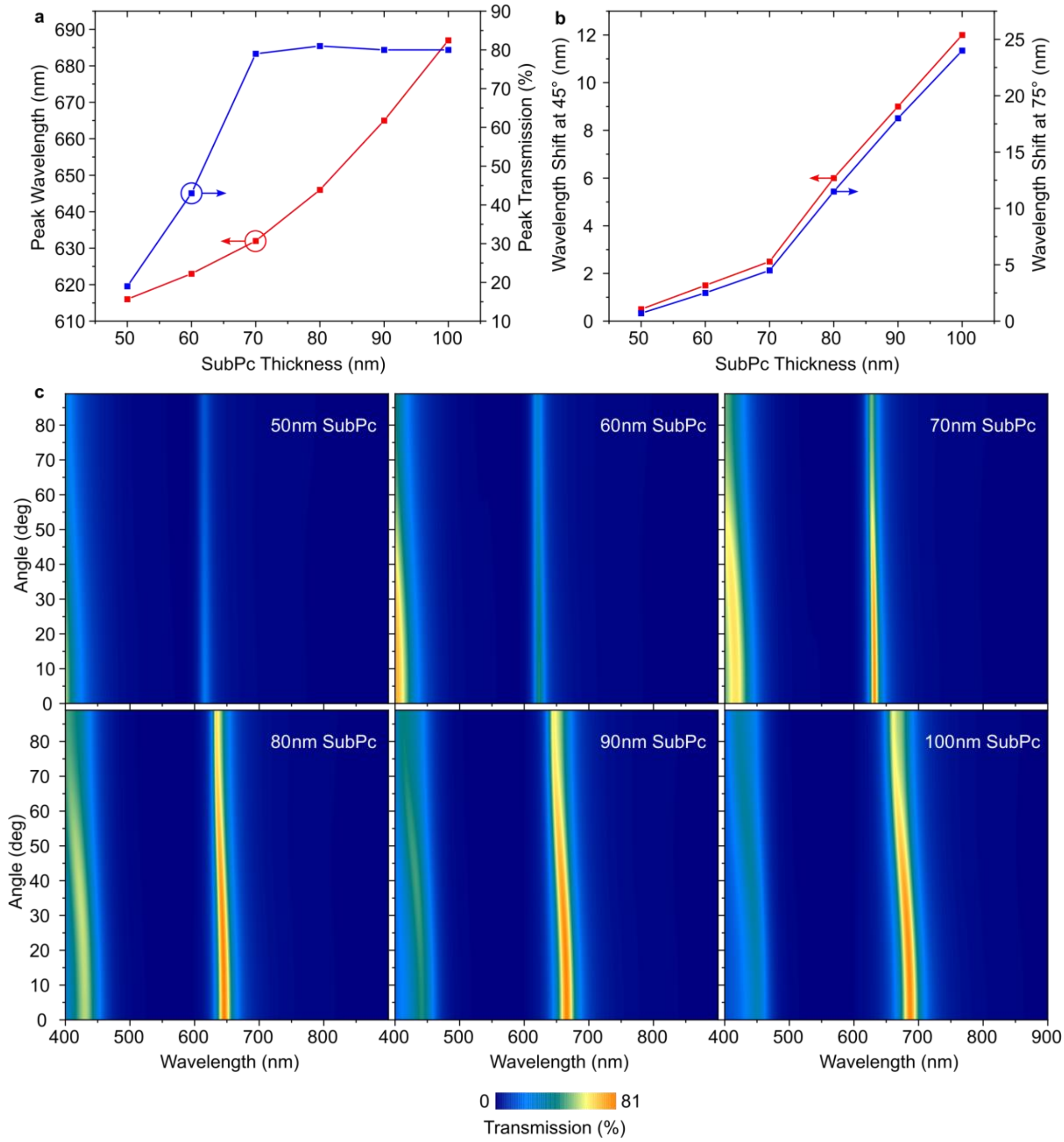

**Figure S13 Tunability of polariton filters. a** Peak wavelength and peak transmission of a 25 nm Ag | SubPc | 25nm Ag polariton filter as a function of the thickness of the SubPc core. By merely changing the thickness of the SubPc layer and thus the detuning of the cavity photon with respect to the exciton, the transmission wavelength of the polariton filter can be tuned by >50 nm (and by >80 nm when accepting reduced performance); i.e. a single material can be used to cover a broad spectral range. **b** Shift of peak transmission wavelength at 45° and 75° angle of incidence as a function of SubPc thickness. Increasing the SubPc thickness leads to an increasing blue-shift of the lower polariton with angle. However, even at a thickness of 100 nm, the blueshift remains small in absolute terms (~12 nm at 45°, ~24 nm at 75°). **c** Simulated angle-resolved transmission spectra for the data presented in **a** and **b**.

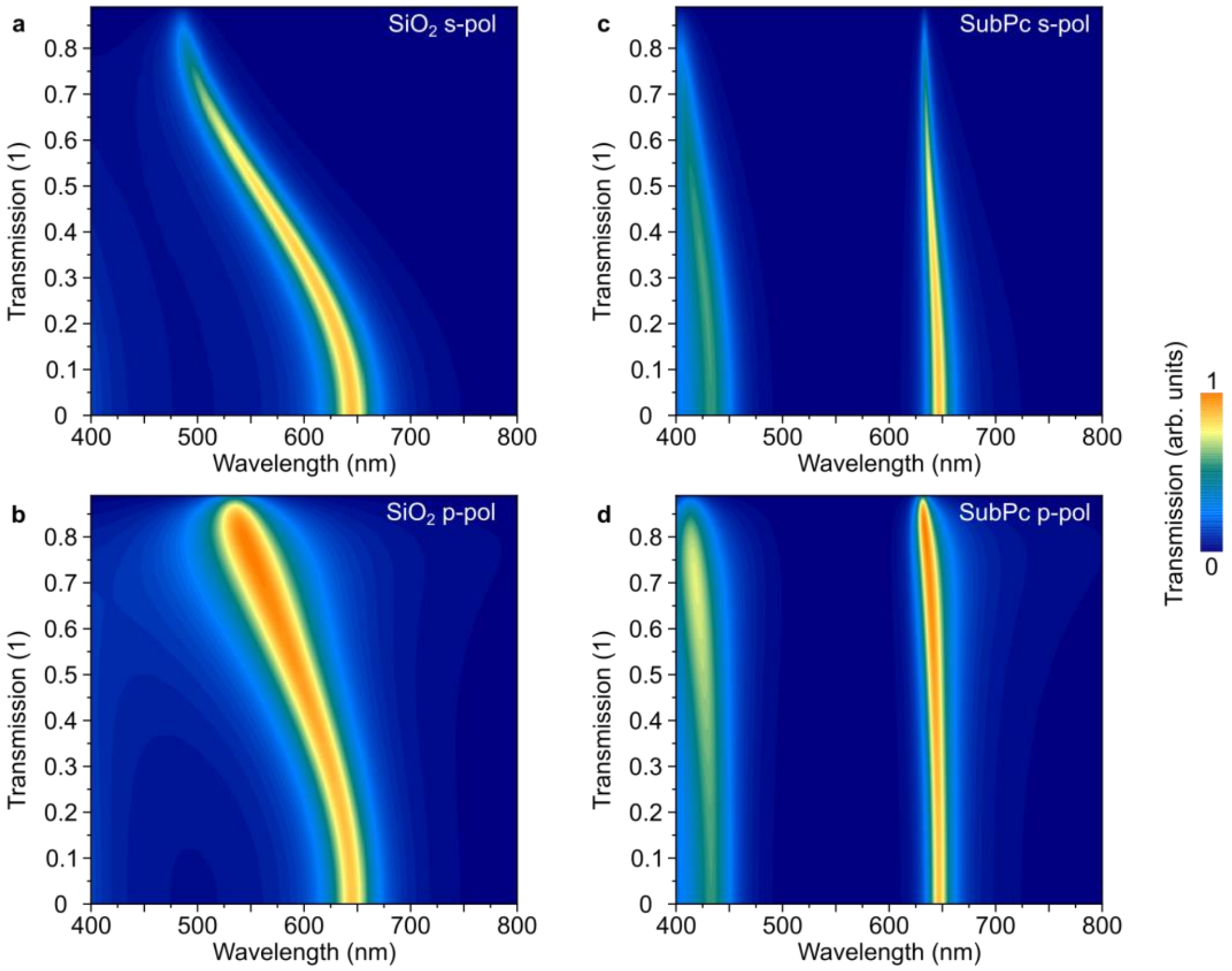

**Figure S14 Polarized transmission of polariton filters.** **a**,**b** Transmission of a 25 nm Ag | 168 nm $SiO_2$ | 25nm Ag MDM filter in s- (**a**) and p-polarization (**b**). **c**,**d** Transmission of a 25 nm Ag | 80 nm SubPc | 25nm Ag polariton filter in s- (**c**) and p-polarization (**d**).

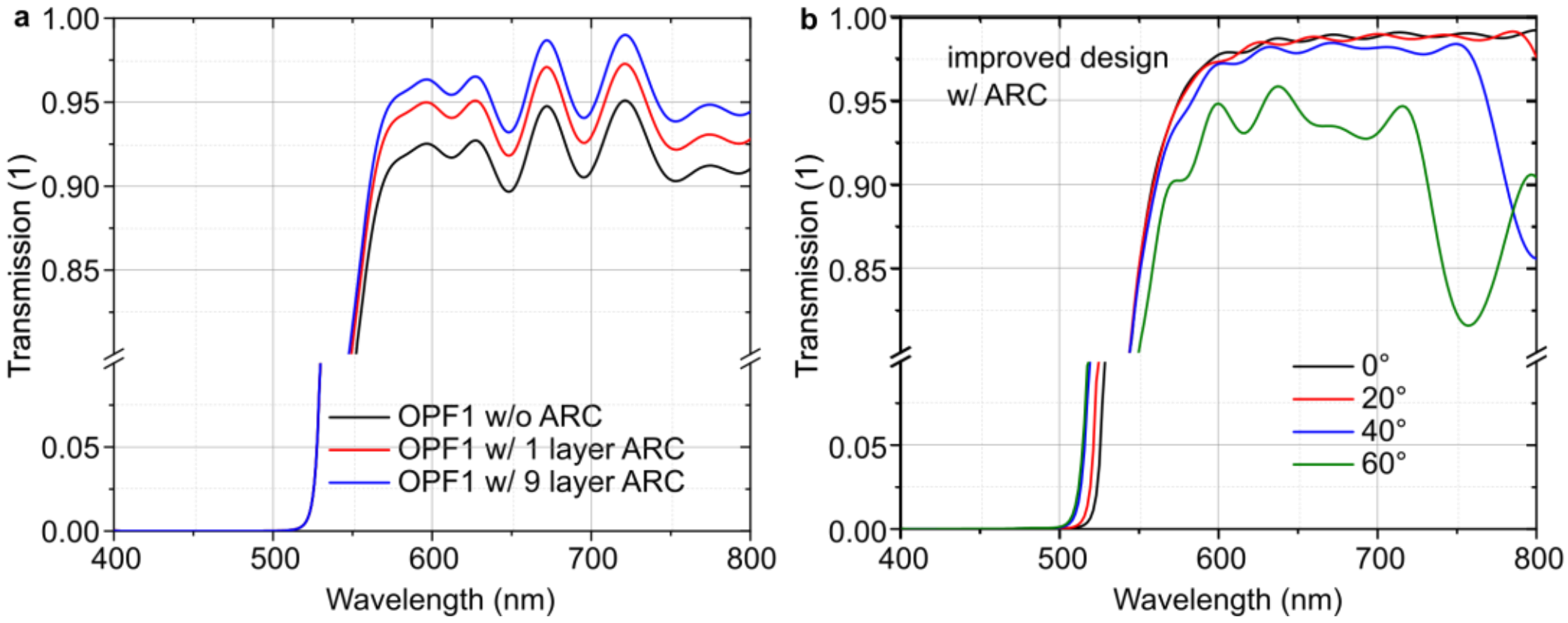

**Figure S15 High-performance design of a DBR-based longpass polariton filter.** **a** Simulated transmission spectra of Optimized Polariton Filter 1 (OPF1) without anti-reflection coating (ARC, black), with a single layer $MgF_2$ backside ARC (red) and with a 9-layer backside ARC (blue). Using a single $MgF_2$ layer as ARC improves transmission by ~3%, using a more complex 9-layer ARC improves transmission by ~5%. **b** Improved design obtained by simultaneous optimization of frontside polariton filter and backside ARC coating. The improved filter shows a stable transmission of >98% for angles up to 40° and >93% up to 60°.

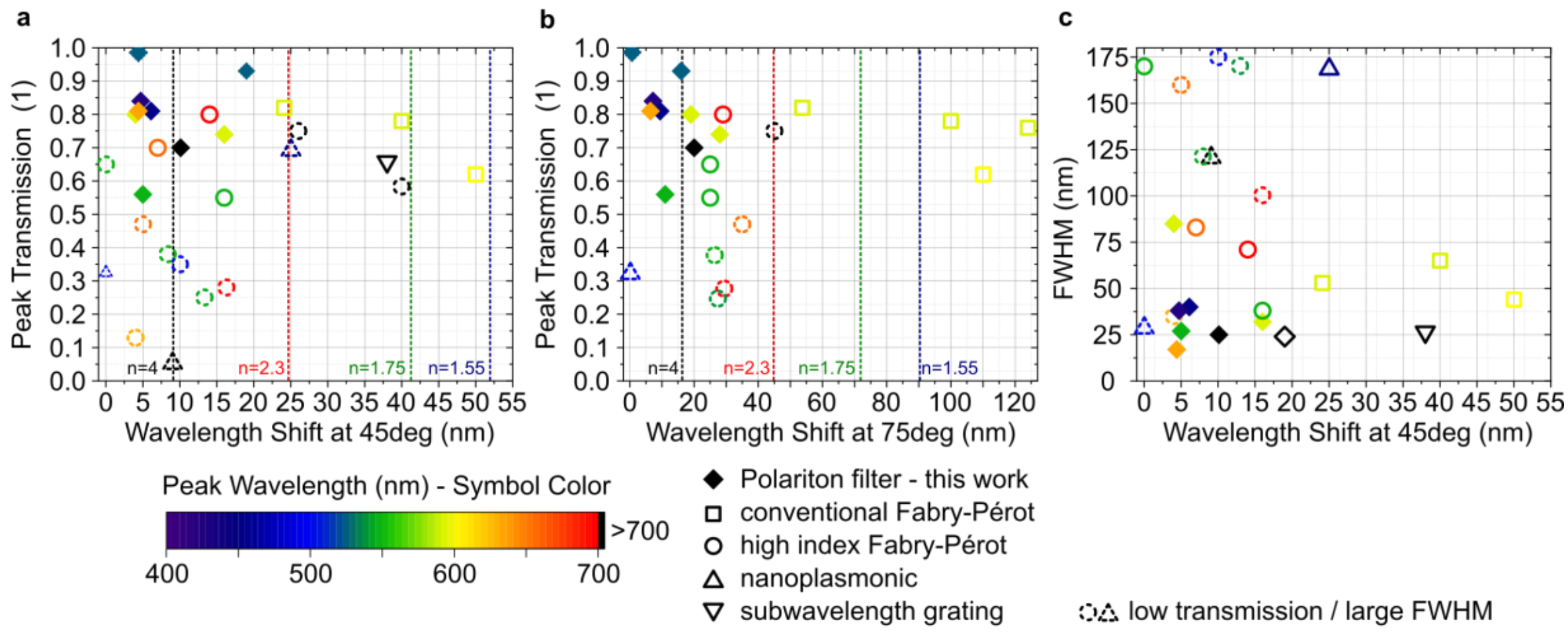

**Figure S16 Literature comparison of angle-stable bandpass transmission filters.** **a** Peak transmission of different filters vs wavelength shift at 45° angle of incidence. **b** Peak transmission of different filter vs wavelength shift at 75° angle of incidence. **c** Full width at half maximum of the main transmission line of different filters vs wavelength shift at 45° angle of incidence. Symbols are coded by colour (peak wavelength) and shape as indicated in the legend. Polariton filters are represented by filled symbols, all other filters are represented by open symbols.

**Table S2: Literature comparison of angle-stable bandpass transmission filters.**

| Filter-Type | Max T[1] | FWHM[2] | Δλ [3] at 45° | Δλ at 75° | λ [4] at 0° | complexity | Ref. |
|---|---|---|---|---|---|---|---|
| Unit | % | nm | nm | nm | nm | | |
| **Polariton Filter - Experiment** | | | | | | | |
| C545T | 56 | 27 | 5 | 11 | 545 | low | this |
| C545T BP | 80 | 85 | 4 | 19 | 590 | low-med | this |
| OPF1 | 93 | LP | 19 | 16 | 525 | medium | this |
| **Polariton Filter - Simulation** | | | | | | | |
| C545T | 74 | 32 | 16 | 28 | 590 | low | this |
| SubNc | 70 | 25 | 10.1 | 20 | 802 | low | this |
| SubPc | 81 | 17 | 4.4 | 6.4 | 638 | low | this |
| BSBCz | 81 | 40 | 6.1 | 9.4 | 452 | low | this |
| SpiroTTB | 84 | 38 | 4.7 | 7.2 | 423 | low | this |
| OPF1-ARC | 98 | LP | 4.4 | 1.1 | 530 | medium | this |
| **Conventional FP filter - Simulation** | | | | | | | |
| SiO2 | 76 | 45 | 61 | 124 | 590 | low | this |
| Ta2O5 | 82 | 53 | 24.1 | 53.7 | 590 | low | this |
| **Literature on angle-stable bandpass transmission filters** | | | | | | | |
| FP cavity | 62 | 44 | 50 | 110 | 600 | low | 4 |
| FP cavity | 78 | 65 | 40 | >100 | 600 | low | 5 |
| HI-FP cavity | 55 | 38 | 16 | >25 | 550 | medium | 6 |
| HI-FP cavity | 75 | 190 | 26 | 45 | 950 | medium | 7 |
| HI-FP cavity | 70 | 83 | 7 | - | 660 | medium | 8 |
| HI-FP cavity | 35 | 175 | 10 | - | 500 | medium | 9 |
| HI-FP cavity | 65 | 170 | 0 | 25 | 550 | medium | 10 |
| HI-FP cavity | 25 | 170 | 13 | 27 | 540 | medium | 11 |
| HI-FP cavity | 28 | 100 | 16 | 29 | 720 | medium | 11 |
| HI-FP cavity | 38 | 121 | 8 | 26 | 545 | medium | 12 |
| HI-FP cavity | 58 | 200 | 40 | - | 900 | medium | 13 |
| HI-lossy FP | 47 | 160 | 5 | 35 | 650 | low | 14 |
| HI-DBR-metal | 13 | 35 | 4 | - | 635 | high | 15 |
| HI-DBR | 80 | 71 | 14 | 29 | 700 | med-high | 16 |
| Polaritonic cavity | low | 24 | 19 | >35 | 900 | low | 17 |
| DBR-plasmonic | 33 | 30 | 0 | 0 | 500 | high | 18 |
| Nanorod | 6 | 122 | 9 | - | 1400 | high | 19 |
| Plasmonic array | 70 | 170 | 25 | - | 450 | high | 20 |
| DBR-grating | 65 | 25 | 38 | - | 800 | high | 21 |

[1] maximum transmission of the main transmission line
[2] full width at half maximum of the main transmission line
[3] wavelength shift of the main transmission line at indicated angle of incidence
[4] central wavelength of the main transmission line

Abbreviations:

OPF1 … Optimized Polariton Filter 1
LP … Longpass
FP … Fabry-Pérot
HI … high index of refraction
DBR … distributed Bragg reflector
\- … data not provided in publication

**Supplementary Discussion 4**

The monolithic filter-photodiode stack is based on two stacked coupled cavities, namely the filter and the diode, with a shared central mirror/anode. Typically, such systems would need to be optically decoupled to avoid interaction between these cavities. Recently, such an organic platform was demonstrated by fabricating filter and diode layers on opposite sides of a mm-thick glass substrate[22]. In the case of polariton filters however, the strong absorption present in the filter cavity efficiently suppresses the formation of coupled optical resonances if all layer thicknesses are optimized.

The current density-voltage (jV) measurements, depicted in Fig. S17, confirm robust diode operation both with and without the additional filter, showing a slightly increased reverse-bias current density for the filtered diode with respect to the reference, indicating that the filter might introduce additional shunts. For both the filtered and unfiltered diodes, we estimate specific detectivities of up to the order of $10^{12}$ Jones, in agreement with previous results on similar diodes[23,24]. To estimate detectivity, we took into account the measured peak external quantum efficiency (EQE) of 18% and describe the electrical noise as sum of thermal and shot noise[25] at zero bias voltage (a full determination of the noise is beyond the scope of the current study).

To accurately model the absorption in the combined filter-photodiode stack, we perform electric field simulations of the device (Fig. S18**a**). The polariton formation in the metal-C545T-metal cavity efficiently filters the light entering the active layer of the photodiode (red shaded area). Using an additional C545T absorption filter can further enhance the spectral selectivity and reduce unwanted signal between 400 nm and 500 nm (Fig. S18**b**).

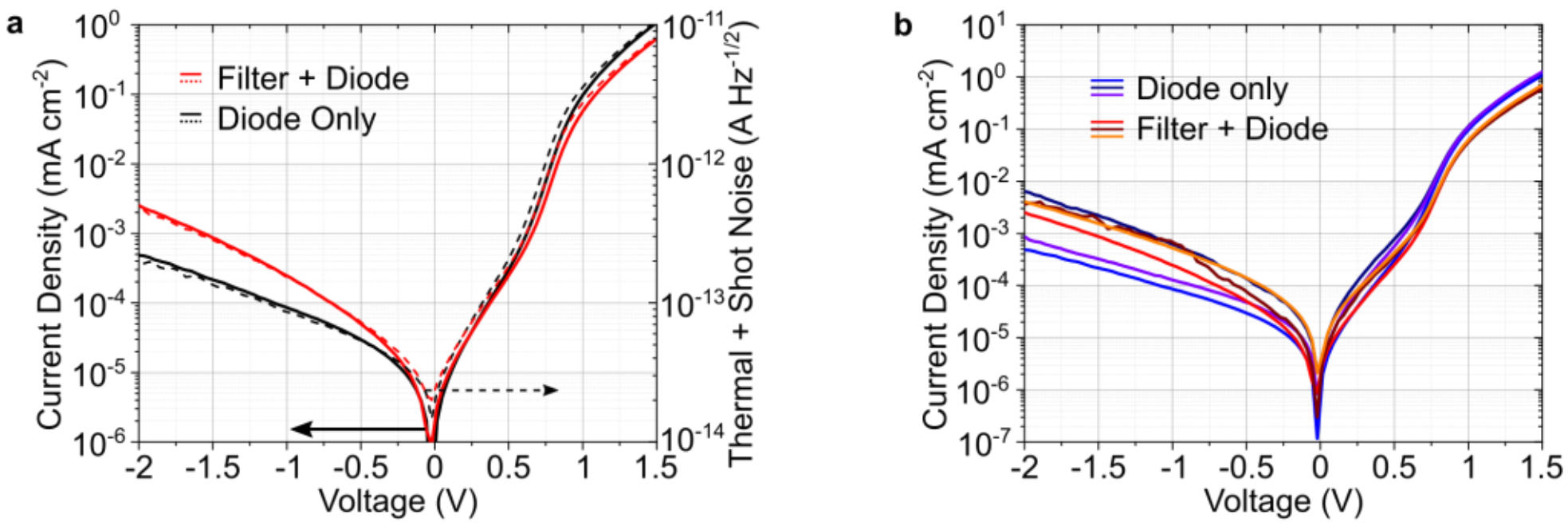

**Figure S17: Current-density – voltage characteristics of filtered and reference photodiodes.** **a** Current density – voltage (solid lines) behaviour and estimated thermal and shot noise current (dashed lines) for the filtered (red lines) and unfiltered (black lines) photodiode. The noise current at 0 V is not influenced by the addition of the polariton filter. **b** Current-density - voltage characteristics for three reference photodiodes and three photodiodes comprising a polariton filter, respectively. The diode characteristics are reproducible for different samples, with the filtered diodes showing a slightly increased reverse current, likely due to higher roughness.

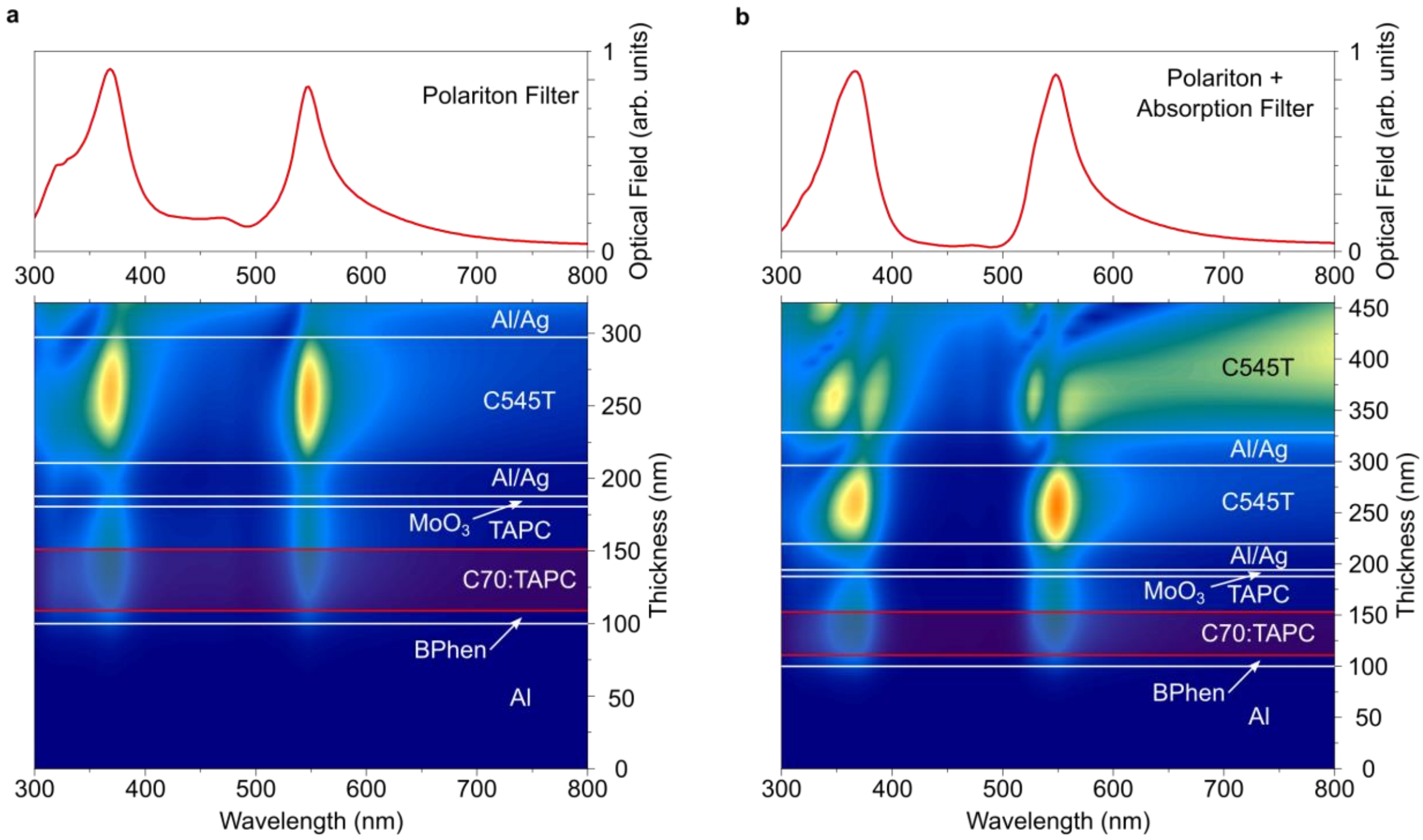

**Figure S18: Simulations of optical field in polariton-filter-photodiode. a** Simulated electric field of the photodiode comprising a polariton filter with the structure 1 nm Al | 25 nm Ag | 80 nm C545T | 1 nm Al | 25 nm Ag | 7 nm $MoO_3$ | 30 nm TAPC | 40 nm C70:TAPC (5wt%) | 12 nm BPhen | 100 nm Al. The profile on top shows the integrated electric field in the photoactive area of the device (C70:TAPC, region marked in red). **b** Simulated electric field of the filtered diode shown in **a** with an additional 150 nm C545T absorption layer on top. This additional absorptive filter reduces the diode signal in the blocking region between 400 nm and 500 nm.

## Supplementary References